\documentclass[%
 preprint, amsmath, amssymb,
 aps, prfluids, superscriptaddress
]{revtex4-2}

\usepackage{graphicx}
 \usepackage{multirow} 
\usepackage{hyperref}
\usepackage{xcolor}

\newcommand{\A}{\text{A}}
\newcommand{\B}{\text{B}}
\newcommand{\C}{\text{C}}
\newcommand{\cA}{c_{\text{A}}}
\newcommand{\cB}{c_{\text{B}}}
\newcommand{\cC }{c_{\text{C}}}
\newcommand{\cAC}{c_{\text{A}+\text{C}}}
\newcommand{\ci}{c_{\text{i}}} 
\newcommand{\isp}{\text{i}} 
\newcommand{\cAt}{\tilde{c}_{\text{A}}}
\newcommand{\cBt}{\tilde{c}_{\text{B}}}
\newcommand{\cCt }{\tilde{c}_{\text{C}}}
\newcommand{\slk}{\ensuremath{\sigma_{\log k}^{2}}}
\newcommand{\lx}{\ensuremath{\lambda_{x}}}
\newcommand{\lz}{\ensuremath{\lambda_{z} }}
\begin{document}

\title{Effect of permeability heterogeneity on reactive convective dissolution}
\author{R. Benhammadi}
\affiliation{Institute of Environmental Assessment and Water Research, Spanish National Research Council (IDAEA-CSIC), Barcelona, Spain}
\author{A. De Wit}
\affiliation{Université Libre de Bruxelles (ULB), Nonlinear Physical Chemistry Unit, CP231, 1050 Brussels, Belgium}
\author{J.J. Hidalgo}
\email{juanj.hidalgo@idaea.csic.es}
\affiliation{Institute of Environmental Assessment and Water Research, Spanish National Research Council (IDAEA-CSIC), Barcelona, Spain}
\date{\today}

\begin{abstract}
The impact of permeability heterogeneity on reactive buoyancy-driven convective dissolution is analyzed numerically in the case of a bimolecular A+B$\to$C reaction across varying Rayleigh numbers. The convective dynamics is compared in homogeneous, horizontally stratified, vertically stratified, and log-normally distributed permeability fields. Key variables, such as the total amount of  product, mixing length, front position and width, reaction and scalar dissipation rates, and dissolution fluxes, are strongly influenced by the type of permeability heterogeneity. Vertically stratified and log-normally distributed permeability fields lead to larger values for all parameters compared to homogeneous fields. Horizontally stratified fields act as an obstacle to convective flow, resulting in slower front progression, thicker fingers, wider reaction fronts, and the lowest dissolution fluxes among all cases.  When the reaction stabilizes convection, flow stagnation occurs near the extremum of the non-monotonic density profile, even in vertically stratified systems, highlighting the complex interaction between reactions and dissolution-driven convection. In log-normally distributed fields, the flow behavior depends on the permeability structure: smaller horizontal correlation lengths cause fingers to spread more horizontally, while larger horizontal correlation lengths promote more vertical movement with shorter wavelengths. Overall, a shorter horizontal correlation length relative to the vertical one leads to an increase in the value of all aforementioned parameters and thus to a more efficient mixing. These findings reveal how heterogeneity affects convective dynamics by influencing the reaction front, dissolution rates, mixing behavior, and mass transport efficiency, emphasizing the intricate role of permeability structure in reactive convective processes. 

\end{abstract}


\maketitle
%
%
\section{Introduction}
Reactive convective mixing, in which both convection and chemical reactions simultaneously influence the transport and mixing of species within a fluid, plays a fundamental role in many environmental and industrial processes. A key application in which such a mixing is important is  the convective dissolution of CO$_{2}$ in saline aquifers, an important process for carbon capture and storage \citep{Kalam2020}. Convective dissolution occurs when CO$_{2}$, injected in deep storage sites,  dissolves in the reservoir brine. As CO$_2$ increases  the fluid's density, it creates an upper denser layer that destabilizes because of a Rayleigh-Taylor instability. This induces convective fingers sinking downwards  \cite{article_Emami, ching_convective_2017,hid13}. This instability helps reducing the risk of CO$_{2}$ leakage, increasing the flux of CO$_2$ towards the host phase and enhancing the long-term stability of storage. Understanding how chemical reactions and heterogeneities of the porous matrix interact with these convective flows is essential for accurately modeling the long-term behavior of CO$_{2}$ in underground reservoirs.

Heterogeneity in geological media, such as variations in permeability and porosity, substantially alters the dynamics of convective dissolution. Real-world CO$_{2}$ reservoirs are far from uniform \citep{Kalam2020}, and previous works on non reactive systems showed that heterogeneity can either accelerate or decelerate convection, depending on the spatial variability of permeability \citep{chen2013, article_Green}. For instance, high-permeability regions tend to initiate convection more quickly, while in vertically stratified media, convection may either be delayed or enhanced depending on the interaction between permeability variations and perturbations \cite{Ghorbani2017}. Heterogeneities also affect the spatial distribution of CO$_{2}$, leading to different flow regimes, such as fingering and channeling, which are governed by factors like correlation length and anisotropy \cite{article_Farajzadeh, ranganathan2012}. Moreover, heterogeneity increases competition between convective fingers, which can reduce the efficiency of dissolution by diminishing the interfacial mixing length \cite{Li2019}. It has also been shown both experimentally and numerically that  heterogeneities in Hele-Shaw cells aperture amplify the characteristics of the instability such as the amplitude and the growth rate \cite{Benhammadi2025}. Similar effects on the growth rate were observed for granular media \cite{article_Brouzet}. Convection has been also investigated in a layered porous medium, focusing on how variations in permeability affect the development of convective patterns \cite{Hewitt_2022}.

In parallel, chemical reactions can as well significantly alter convective  dynamics by modifying the fluid's density \citep{Almarcha2009}.
For convective dissolution, chemical reactions can either accelerate or delay convection depending on whether the reaction products are denser or less dense than the reactants \cite{ghe11,Loodts2014,car14,Loodts2016,Loodts2017,Ghoshal2018,Jotkar2019b,lei19,Tho20,Jotkar2020,Jotkar2021}. Differential diffusivities of the species involved in the reactions can dramatically alter the convection patterns \cite{Jotkar2020,Jotkar2021} and dissolution fluxes can be strongly enhanced by reactions \cite{Loodts2017}. Density profiles can be classified in reactive convective dissolution systems based on the relative solutal density contributions of the reaction product and reactants and the initial concentration ratio of the reactants \cite{Loodts2015,Loodts2016}. By varying these parameters, four distinct convective regimes have been identified, establishing a predictive parameter space that links reaction-driven density changes to the onset and nature of convective instabilities. 

The combined effects of heterogeneity and chemical reactions on reactive convective mixing remain however largely unexplored \cite{lei19}.
In this context, we  investigate here the  impact of combined heterogeneities and chemical reactions on convective density-driven dissolution. Specifically, we examine how variations by reactions in density profiles, both monotonic and non-monotonic, influence the convective instability in heterogeneous porous media. Heterogeneity is introduced through simulations involving both horizontally and vertically stratified porous media, as well as more complex log-normally distributed permeability fields with varying correlation length, anisotropy ratio and variance. This approach enables us to model more realistic geological conditions, offering insights crucial for understanding CO$_{2}$ sequestration in natural reservoirs and advancing the field of reactive convective mixing.

The manuscript is organized as follows: Section 2 introduces the model studied while Section 3 presents the simulation results and discussions. Section 4 concludes with a summary of key findings and suggestions for future research.
%
%
\section{Methodology}
We study convective dissolution of species $\A$ in a host phase containing a species $\B$ in presence of a bimolecular reaction $\A +\B \rightarrow \C$ in the case of homogeneous and heterogeneous porous media. Simulations were performed by adding chemical reactions to the variable-density flow and transport solver \texttt{rhoDarcyFoam} belonging to the open-source computational framework SECUReFoam \citep{Icardi2023}, based on the finite-volume library OpenFOAM \citep{Weller1998}.
%
%
\subsection{Governing equations}
Assuming fluid incompressibility, the dimensionless governing equations of fluid flow, solute transport and reaction in a porous medium are \citep{Loodts2017}
\begin{align}  
\label{eq:continuity}
\nabla \cdot \mathbf{q} = 0   
\end{align}
\begin{align} 
\label{eq:RDCA}
\frac{\partial \cA}{\partial t} + \mathbf{q} \cdot \nabla \cA = \nabla^2 \cA - \cA\cB
\end{align}
\begin{align}  
\label{eq:RDCB}
  \frac{\partial \cB}{\partial t} + \mathbf{q} \cdot  \nabla \cB = \delta_{\B}\nabla^2\cB - \cA\cB
\end{align}
\begin{align}  
\label{eq:RDCC}
\frac{\partial \cC}{\partial t} + \mathbf{q} \cdot  \nabla \cC = \delta_{\C} \nabla^2 \cC + \cA\cB,
\end{align}
where $c_{\isp}$ and $D_{\isp}$ are the dimensionless concentration and diffusion coefficient of specie $\isp$ $(\isp = \A, \B, \C)$, $\delta_{\B}=\frac{D_{\B}}{D_{\A}}$, $\delta_{\C}=\frac{D_{\C}}{D_{\A}}$,  and $\mathbf{q}$ is the dimensionless Darcy's velocity
\begin{align}  
\label{eq:Darcy}
   \mathbf{q} = -k(\nabla p - \rho \mathbf{e}_{z})    
\end{align}
with $k$ the porous medium  permeability, $p$ the  pressure, $\rho$ the  fluid density and $\mathbf{e}_{z}$ a unit vector in the direction of gravity.

The above set of equations is obtained by scaling concentration by the solubility $c_{\A0}$ of species $\A$ and taking the characteristic time, length and velocity as $t_{c} = \phi/\kappa c_{\A 0}$,  $\ell_{c} = \sqrt{D_{\A} t_{c}}$ and $q_{c} = \phi \ell_{c}/t_{c}$ respectively, where $\phi$ is porosity and $\kappa$ the kinetic constant of the reaction.

The dimensional fluid density is assumed to be a linear function of the dimensional species concentration $\tilde{c}_{\text{i}}$
\begin{align}
  \label{eq:rho_fluid}
  \tilde{\rho} = \rho_{0} \left( 1 + \alpha_{\A} \cAt + \alpha_{\B} \cBt + \alpha_{\C} \cCt \right),
\end{align}
with $\rho_{0}$ the density of the solvent and $\alpha_{\text{i}} = \frac{1}{\rho_{0}}\frac{\partial \tilde{\rho}}{\partial \tilde{c}_{\text{i}}}$  the solutal expansion coefficient of species $\text{i}$. The dimensionless density is defined as
\begin{align}
  \label{eq:rho_def}
  \rho = \frac{\tilde{\rho} - \rho_{0}}{\rho_{c}} = R_{\A}\cA + R_{\B}\cB + R_{\C}\cC,
\end{align}
where $\rho_{c} = \phi \mu D_{\A}/(gk_{c}\ell_{c})$ is the density scale, $\mu$ is the fluid viscosity, $g$ the gravity acceleration, $k_{c}$ is the geometric mean of the permeability field and
\begin{align}
  \label{eq:Ri}
  R_{\isp} = \frac{\alpha_{\text{i}} c_{\A 0} \rho_{0} g k \ell_{c}}{\phi \mu D_{\A}}
\end{align}
are the species Rayleigh numbers. They quantify the contribution of the species $\isp$ to the dimensionless density. The sign of $R_{\isp}$ indicates whether a species increases ($R_{\isp} > 0$) or decreases ($R_{\isp} < 0$) the density of the solution \cite{Loodts2016}.

Equations \eqref{eq:continuity} -- \eqref{eq:RDCC} are solved in a rectangular domain of width $L=3072$ and height $H=2048$ in which the species $\A$ dissolves from above and reacts with $\B$ to produce $\C$. We impose periodic boundary conditions on the left and right boundaries and no-flow boundary conditions on the rest of the boundaries. The concentration of species $\A$ is prescribed at the top boundary i.e. we take
\begin{align}
  \label{eq:BCq}
  \mathbf{q}(0, z, t) &= \mathbf{q}(L,z, t); && \mathbf{q} \cdot \mathbf{n} = 0 \text{ at } z=0,H;
\end{align}
\begin{align}
  \label{eq:BC_periodic}
  \ci(0, z, t) &= \ci(L, z, t), \text{ i} = \A, \B, \C; 
\end{align}                                     
\begin{align}
  \label{eq:BCbc}
&& \nabla \ci \cdot \mathbf{n} = 0 \text{ at } z=0, H, \text{ i} =\B, \C;
\end{align}
\begin{align}
  \label{eq:BCa}
  \cA(x, 0, t)= 1;
\end{align}

Initially, the domain is filled with species $\B$ and concentration of species $\A$ is perturbed at the top boundary to trigger the instability. Thus
\begin{align}  
\label{eq:IC0}
  \cA(x, z, 0) =
    \begin{cases}
      1 + \epsilon\,\xi(x) & \text{if } z=0\\
      0 & \text{otherwise}
    \end{cases}       
\end{align}
\begin{align}  
\label{eq:IC}
  \cB(x, z, 0) &= \beta \\
  \cC(x, z, 0) &= 0,
\end{align}
where $\beta = \frac{\B_{0}}{\A_{0}}$ is the ratio between the initial concentration $\B_{0}$ of species $\B$ and the solubility $\A_{0}$ of $\A$ in the host phase, $\xi(x)$ is a random uniform distributed variable between 0 and 1, and $\epsilon = 10^{-3}$ is the amplitude of the perturbation.

Taking $\delta_{\B}=\delta_{\C} = 1$, in order to avoid double-diffusive instabilities \cite{Jotkar2020}, and under the above boundary and initial conditions, $\cC = \beta - \cB$, obtained by adding \eqref{eq:RDCB} and \eqref{eq:RDCC}. We can then rewrite density \eqref{eq:rho_def} as
\begin{align}
  \label{eq:rho2}
  \rho = R_{\A}\cA + \left(R_{\B} - R_{\C}\right)\cB + \beta R_{\C},
\end{align}
which shows that the density profile depends on $R_{\A}$,  $\left(R_{\B} - R_{\C}\right)$ and $\beta$ \citep{Loodts2014,Loodts2015,Loodts2016}. We consider the reactive cases summarized in Table \ref{tab:reactive_cases}. We choose two cases in which the non reactive system is already genuinely unstable because $R_A>0$ i.e. species A increases the density upon dissolution of A (cases R1 and R2). In the case R1, $\A$ and $\B$ are consumed by the reaction and replaced by a less dense C. This triggers a non-monotonic profile with a minimum stabilizing convection \citep{Loodts2014}. In the case R2, all species increase density and the system becomes more unstable as the reaction consumes species $A$ and $\B$ to replace them by a denser $C$ \citep{Loodts2014}. This triggers the instability earlier compared to the other cases, thus resulting in strong convection \citep{Loodts2017}.  We add a third case R3 for which $R_A<0$ such that the non reactive system is stable but convection can be triggered if the reaction builds up a non-monotonic density profile  \cite{Jotkar2019b,Jotkar2021}. The case R3 is initially stable but the density profile becomes unstable as the reaction progresses and $\cC$ increases locally. 
\begin{table}
  \centering
  \begin{tabular}{|l|c|c|c|c|}
    \hline
    \multicolumn{1}{|c|}{Case} & $\beta$ & $R_{A}$ & R$_{B}$ & R$_{C}$\\
    \hline
    R1 least unstable reactive & 1 & 1 & 1 & 0 \\
    \hline
    R2 most unstable reactive & 1 & 1 & 1 & 2 \\
    \hline
    R3 unstable due to reaction & 1 & -1 & 1 & 2 \\
    \hline
  \end{tabular}
  \caption{Values assigned to the species Rayleigh numbers and $\beta$ for the different cases}
  \label{tab:reactive_cases}
\end{table}
%
%
\subsection{Heterogeneity}
We consider log-normally distributed permeability fields with a variance of log-permeability $\slk= 1,2,3$. Three different spatial permeability structures are considered: horizontally stratified, vertically stratified and multi-Gaussian random fields with a Gaussian autocorrelation function.

\begin{figure}
    \centering
    \includegraphics[width=\linewidth]{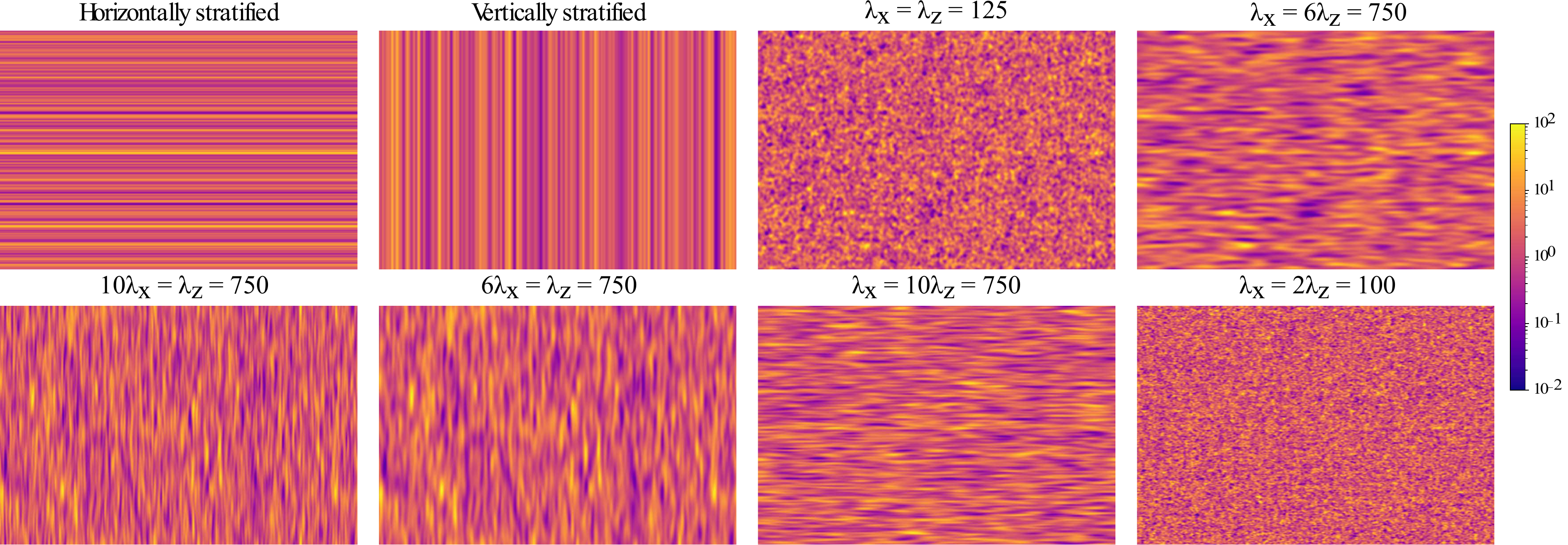}
    \caption{Different realizations of the log-permeability fields for the horizontally stratified,  vertically stratified, and multi-Gaussian permeability structures with variable correlation lengths \lx and \lz. All realizations have $\slk =1$.}
    \label{fig:permeability}
\end{figure}

For the horizontally stratified case, the vertical correlation length \lz{} is set to $2/3$ of the mixing length (i.e., the distance between the tip and rear of the fingers \citep{Wit2004}) observed in the homogeneous R1 case at the onset of finger merging. For the vertically stratified case, the horizontal correlation length \lx{} is taken as the dominant wavelength of the R1 homogeneous case. In the multi-Gaussian case, \lx{} is chosen as a multiple of the correlation length at which resonance in mixing length growth occurs in the vertically stratified R1 case~\cite{DeWit1997b}. The vertical correlation length \lz{} is then varied to explore the effect of anisotropy. The values of \lx{} and \lz{} used are summarized in Table~\ref{tab:corr_length}, and examples of the generated permeability fields with a variance $\slk=1$ are shown in Figure~\ref{fig:permeability}.

Each heterogeneous simulation is repeated 10 times with different white noise realizations to reduce variability and ensure statistical robustness. Results are averaged across the runs.

%
\begin{table}
  \centering
  \begin{tabular}{l|c}
    \hline
    Horizontally stratified & $\lx=\infty; \lz = 50$  \\
    \hline
    Vertically stratified & $\lx= 100; \lz = \infty$ \\
    \hline
    \multirow{5}{*}{Multi-Gaussian} & $\lx= \lz = 125$   \\ 
                             & $\lx= 2\lz = 100$
                           \\  & $\lx= 6\lz = 750$   \\ 
                               & $10\lx= \lz = 750$\\ 
                               & $6\lx= \lz = 750$ \\ 
                               & $\lx= 10\lz = 750$ \\
    \hline
  \end{tabular}
  \caption{Horizontal and vertical correlation lengths of the log-permeability fields used for the horizontally stratified, vertically stratified and multi-Gaussian permeability structures.}
  \label{tab:corr_length}
\end{table}

%
%
\section{Results and discussion}
Let us analyze how heterogeneity affects buoyancy-driven reactive convective dissolution for the considered horizontally stratified, vertically stratified and multi-Gaussian cases with and without reactions. Specifically, we analyze how the combined effect of heterogeneities and reaction impacts the convective fingering patterns, the mixing length and dissolution fluxes, the amount of reaction product and the properties of the reaction front.

\subsection{Density patterns}
\begin{figure}
    \centering
\includegraphics[width=0.99\textwidth]{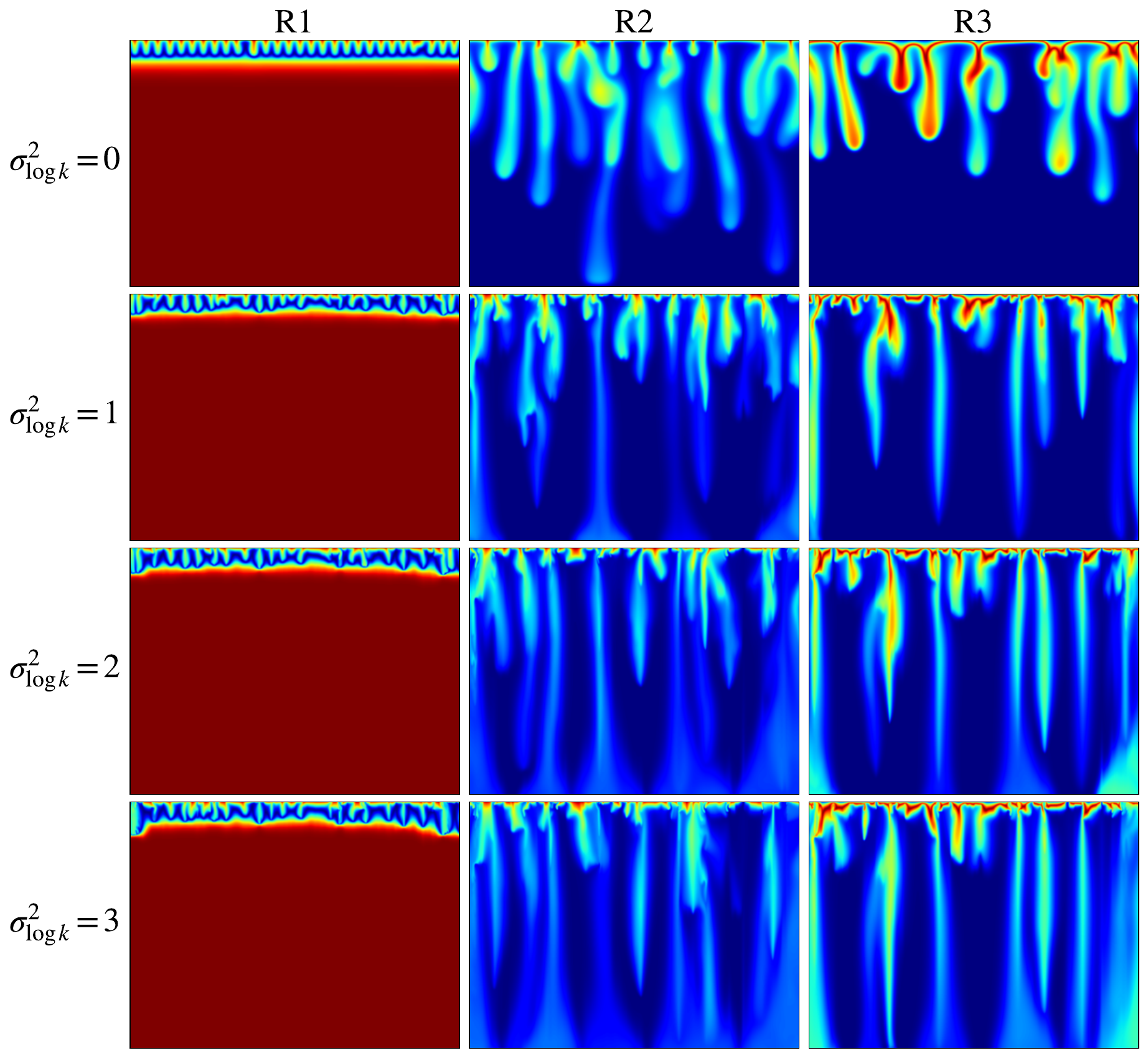}
    \caption{Density patterns for the homogeneous case at time t=8000 and the vertically stratified permeability with different variances at $t=4000$ for R1, R2 and R3 cases. Fingers in the homogeneous case are smoother with rounded tips whereas vertically stratified ones are thinner, elongated and with sharp tips. In R1, the non-monotonic density profile with a minimum  stabilizes convection and the front looks like stuck. In contrast, in the more unstable R2 and R3 cases, the fingers reach the bottom of the domain faster. Increasing $\sigma^2_{\log k}$ makes the fingers reach the bottom of the domain even earlier}
    \label{fig:rhomapvertstrat}
\end{figure}

\begin{figure}
  \centering \includegraphics[width=0.99\textwidth]{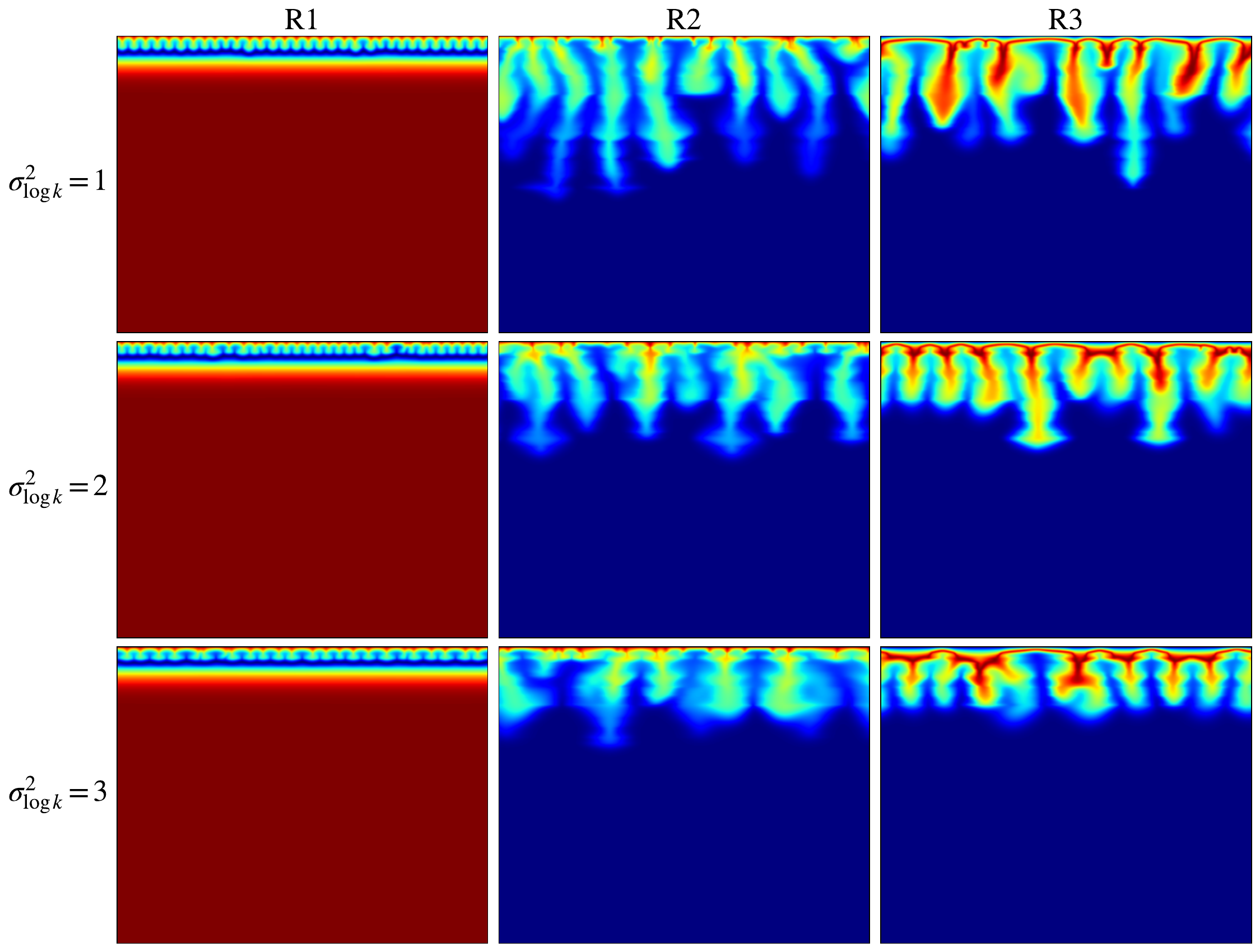}
  \caption{Density patterns for the horizontally stratified permeability with different variances at $t=8000$ for R1, R2 and R3 cases. Fingers are thick and spread laterally. Merging is more pronounced in R2 and R3 cases when $\sigma^2_{\log k}$ is increased. Fingers advance more slowly in R1 as $\sigma^2_{\log k}$ is increased} \label{fig:rhomaphorizstrat}
\end{figure}

Figures~\ref{fig:rhomapvertstrat} and \ref{fig:rhomaphorizstrat} show the fingering patterns obtained for the vertically stratified and horizontally stratified media, respectively. In the homogeneous medium (Fig.\ref{fig:rhomapvertstrat} top), the fingers appear smoother with rounded tips compared to those in the heterogeneous media. The fingers in the vertically stratified case are more elongated, thinner and exhibit sharper tips. They tend to align with regions of high permeability, as observed by \citet{Ghorbani2017}. In contrast, the fingers in the horizontally stratified cases (Fig.\ref{fig:rhomaphorizstrat}) appear thicker, primarily because they tend to spread laterally, as the horizontal stratification acts as a barrier to their vertical movement.
%
%

\begin{figure}
    \centering \includegraphics[width=0.49\linewidth]{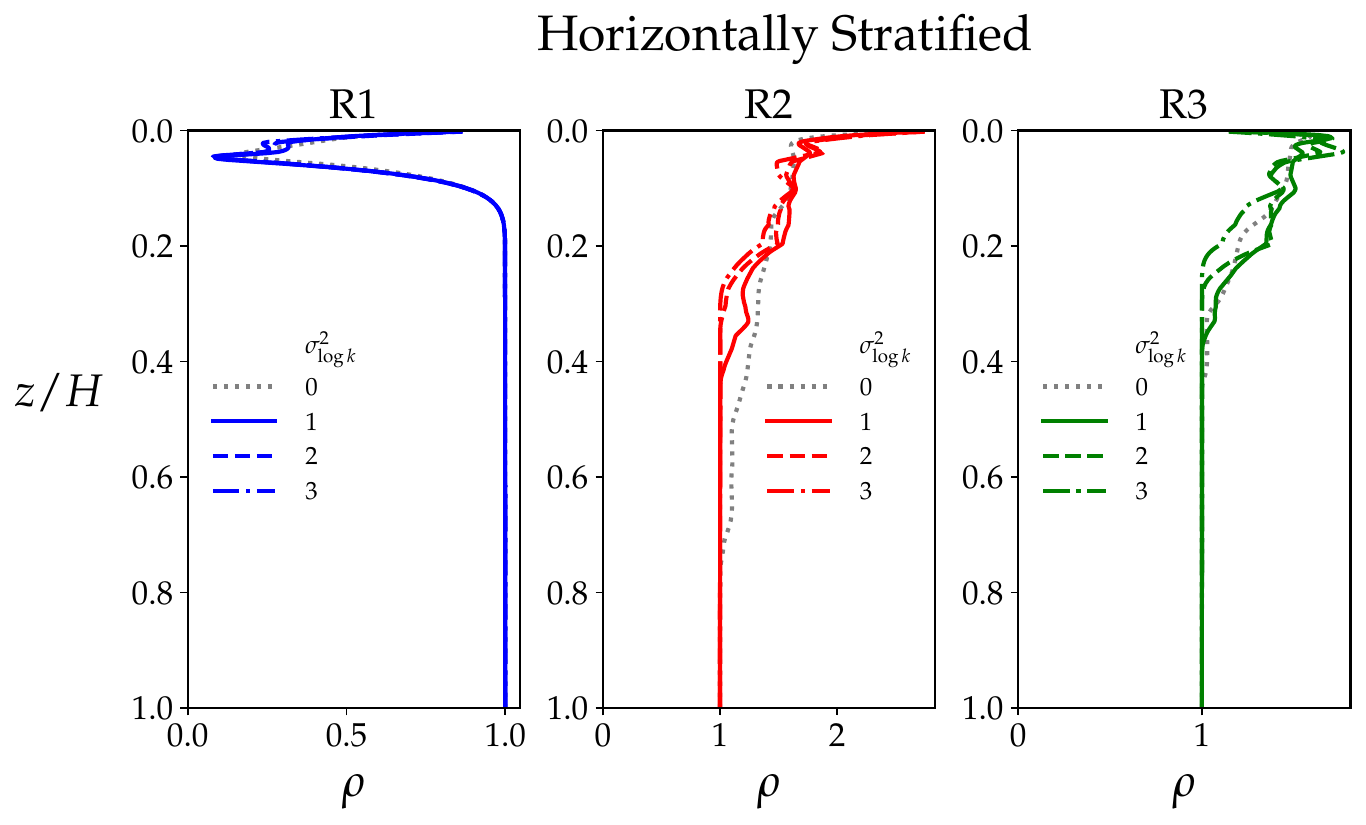} \includegraphics[width=0.49\linewidth]{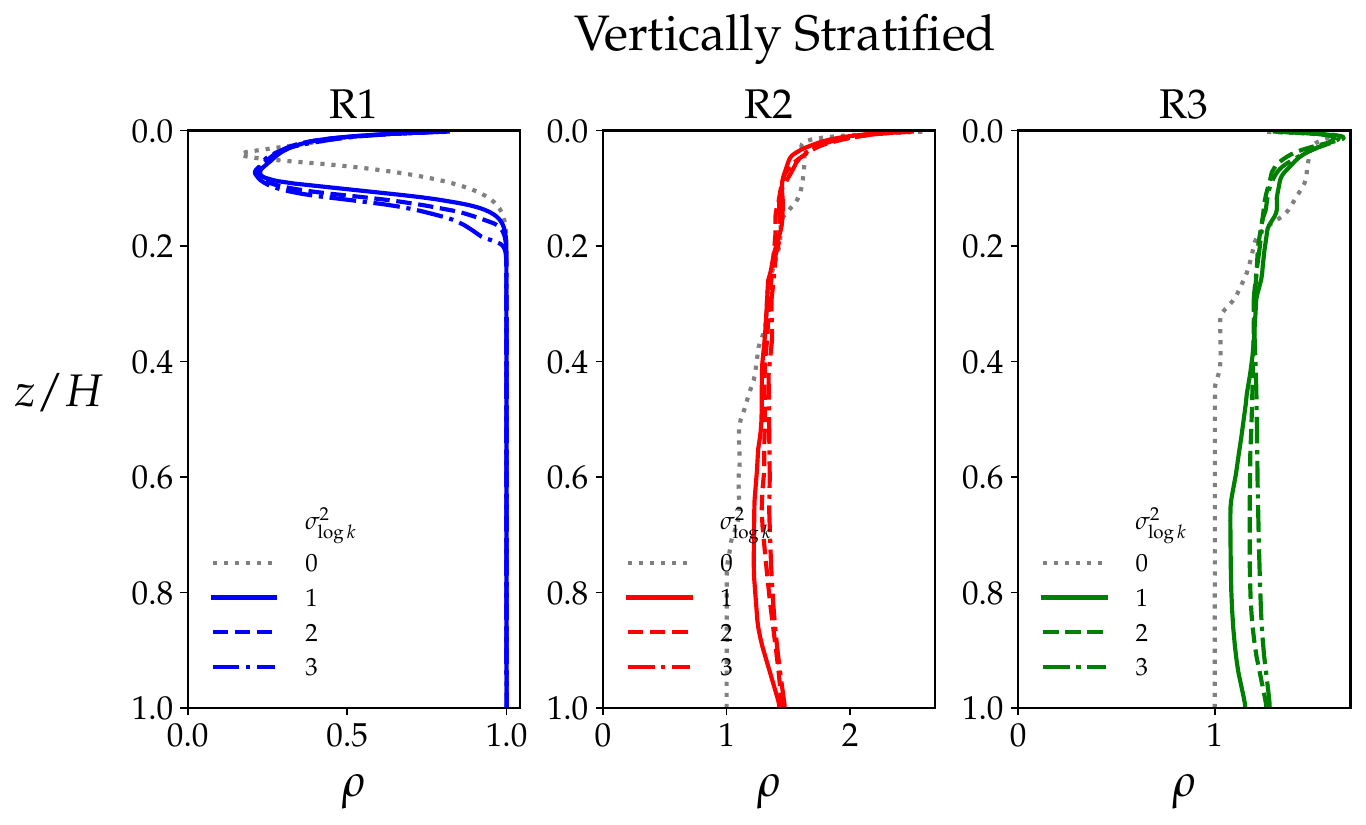}
    \caption{Density profiles of the reactive homogeneous and stratified cases at $t=6000$. R1 shows a non-monotonic density profile whereas R2 and R3 feature a monotonic decreasing density.}
    \label{fig:rho_compstrat}
\end{figure}

In the case R1, convection is confined in an upper layer, even for the heterogeneous cases, because of a minimum in the density profile, as depicted in the non-monotonic density profile shown in Fig.\ref{fig:rho_compstrat}. This behavior aligns with the linear stability analysis  \cite{Loodts2014} and non-linear simulations  \cite{Loodts2017}, which demonstrated that when a non-monotonic density profile with a minimum develops over time, reactions tend to stabilize convection. Case R2 corresponds to reactive systems with monotonic density profiles where convection is accelerated because the reaction product $\C$ is significantly denser than reactant $\B$. Fingers progress then the fastest. This case is considered as the most realistic scenario for CO$_2$ sequestration because of the combined influence of all species contributing to an increase in the density profile, particularly if species $\C$ has the highest density ratio, R$_{\C}=2$ (see Table~\ref{tab:reactive_cases}). In the case R3,  one observes that the fingers move faster than R1 but slower than R2 because the species $\A$ contributes negatively to the density profile. Furthermore, one notices from the density profiles (Fig.\ref{fig:rho_compstrat}) that in the case R1, all density profiles stabilize around $\rho=1$ whereas in cases R2 and R3, the densities spread out more and do not cluster around a single value of $\rho$ because of stronger buoyancy and chemical reactions effects. 

A visual inspection of the fingers for the different reactive cases in the horizontally stratified case (see figure \ref{fig:rhomaphorizstrat}) shows that the merging of the fingers happens earlier when the variance is increased and that the fingers look more distorted than in the homogeneous case. Similarly, the transition to the convective regime tends to happen much earlier in the vertically stratified cases when the variance is increased (see figure \ref{fig:rhomapvertstrat}).

We also observe, as in \citet{Hewitt_2022}, that increasing the variance of horizontally stratified permeability has a similar effect to increasing the impedance, leading to earlier suppression of vertical convective flow. This suppression promotes the formation of localized convective cells, which develop within higher-permeability zones enclosed by low-permeability barriers as shown in figure \ref{fig:rhomaphorizstrat}. 

\begin{figure}
    \centering
\includegraphics[width=0.7\textwidth]{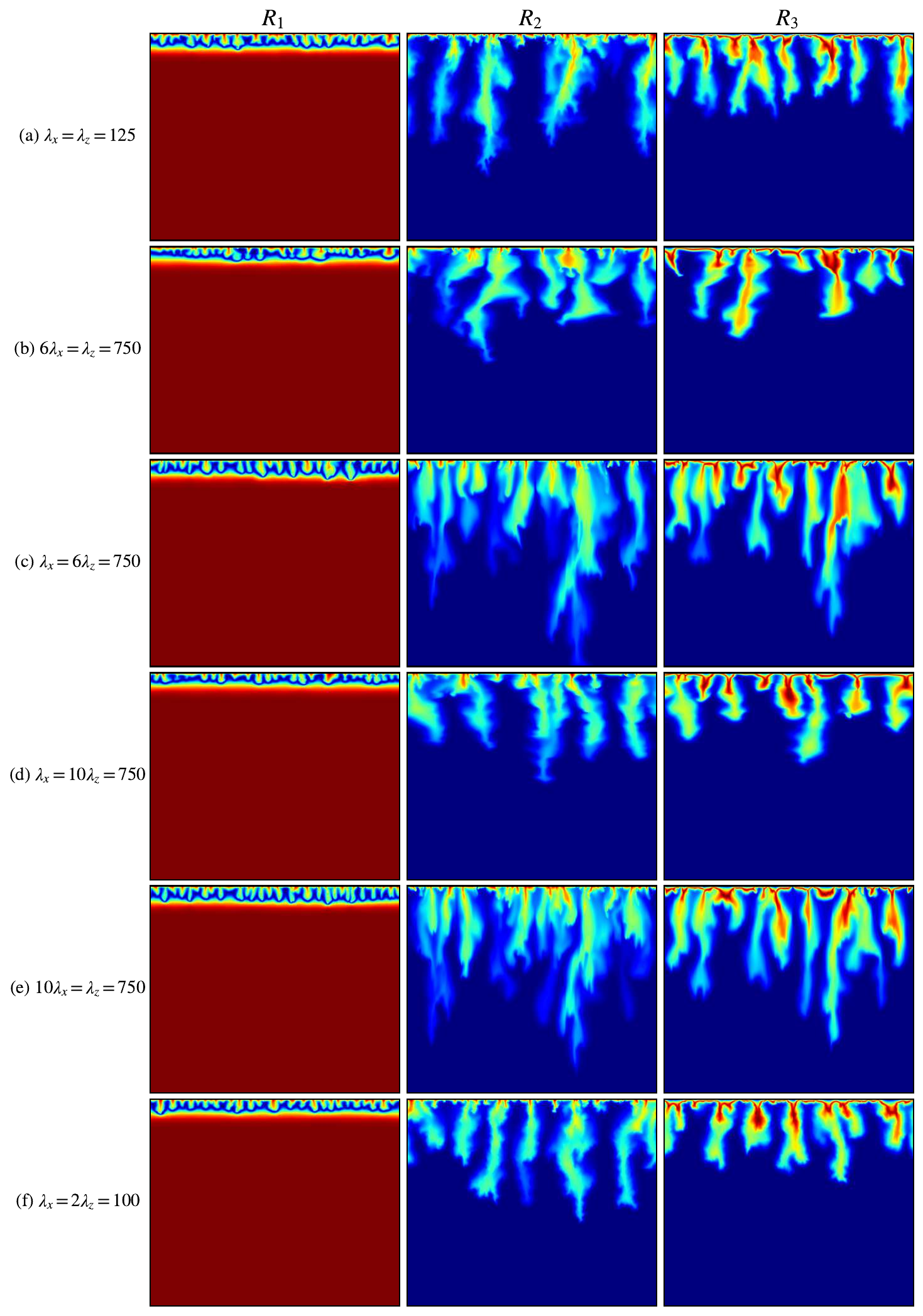} 

    \caption{Log-Normally distributed permeability cases density at $t=4000$ for different correlation lengths and for $\slk=1$. Cases with $\lx<\lz$ ((b) and (d)) present fingers similar in shape with the horizontally stratified ones whereas cases with $\lx>\lz$ ((c) and (e)) present fingers similar in shape with the vertically stratified ones. The isotropic case (a) exhibits fingers that resemble roots or small channels. As previously observed for the stratified cases, increasing \slk{} accelerates convection.}
    \label{fig:rholog4000}
\end{figure}

In multi-Gaussian fields (Figure~\ref{fig:rholog4000}), the fingers predominantly propagate laterally when the anisotropy ratio  $\lx/\lz > 1$ (Fig.\ref{fig:rholog4000}c and d), similarly to the horizontally stratified cases. Their vertical advancement is impeded by the high permeability regions. Additionally, the large anisotropy ratio case (c) shows faster finger merging, a more advanced density front and a smaller number of fingers than in the small anisotropy ratio case (d). Conversely, when $\lx/ \lz < 1$ (Figure~\ref{fig:rholog4000} (b) and (e)), the fingers progress more easily in the vertical direction and are thinner, similarly to the vertically stratified case. The velocity at which the density front advances is proportional to the anisotropy ratio,  case (b) being faster than case (e).

Furthermore, increasing \slk{} makes the front advance more slowly for the $\lx/\lz > 1$ cases, because of the barrier effect of the low permeability zones, and faster for the $\lx/\lz < 1$ as the vertical structures induce stronger channeling.
%
%
%
\subsection{Mixing length}
The mixing length, which characterizes the non-linear dynamics of the fingers, is defined as the distance between the tip and the rear of the fingers \citep{Wit2004}. In practice, it is computed here as the vertical distance over which $\rho - R_{\B} > 10^{-3}$.

\begin{figure}
  \centering \includegraphics[width=0.49\textwidth]{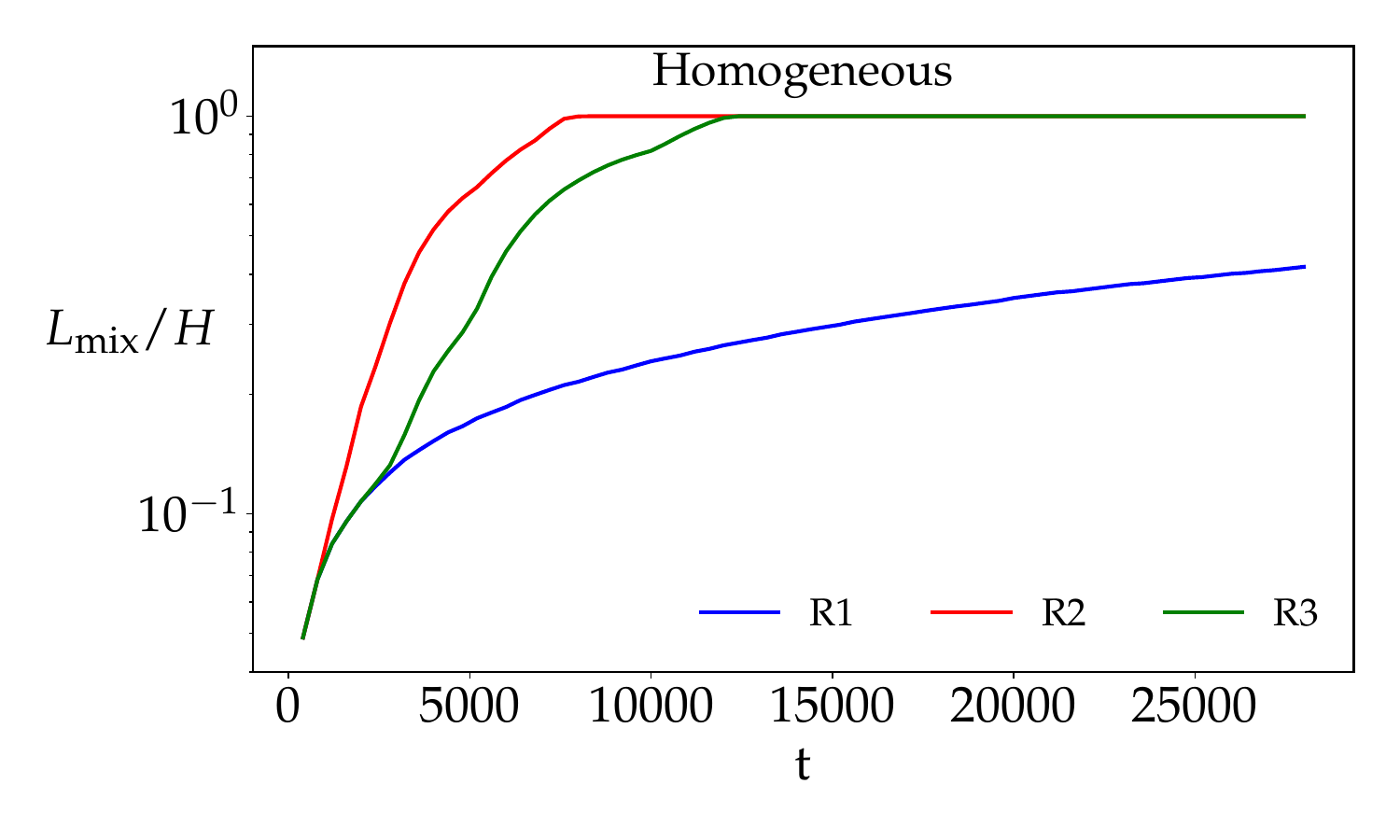}\\
  \includegraphics[width=0.49\textwidth]{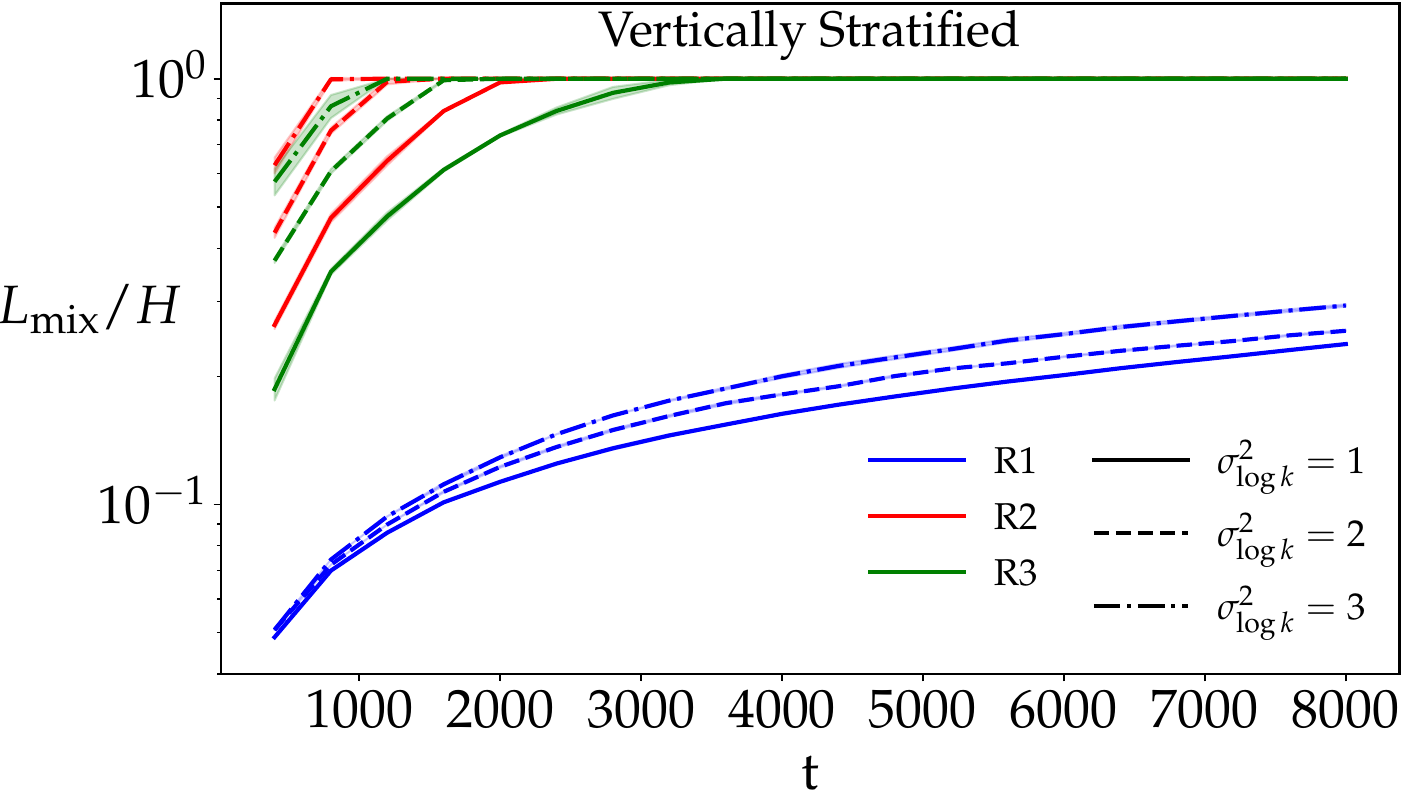}
  \includegraphics[width=0.49\textwidth]{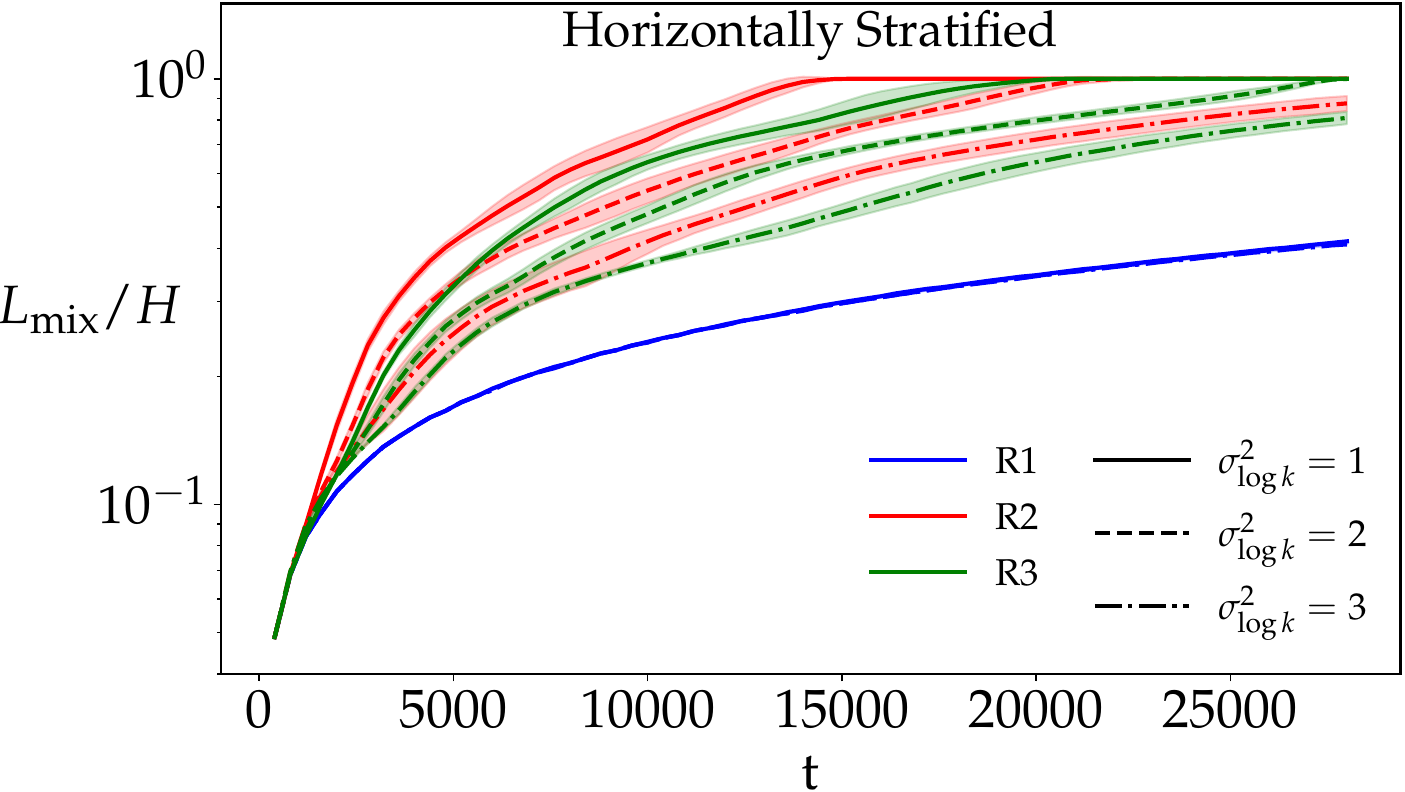}
    \caption{Mixing length for the homogeneous and all stratified cases. Vertically stratified cases present steeper slopes and their mixing lengths increase proportionally to the variance. Horizontally stratified cases display the opposite trend, their mixing lengths decrease with the variance. Notably, in all cases, the diffusive regime is only captured in the least unstable case R1. The shaded area shows the variability between realizations.}
    \label{fig:mixlencompstrats}
\end{figure}

The mixing length for the homogeneous and stratified cases is shown in Figure \ref{fig:mixlencompstrats}. Before the onset of the instability, while diffusion is the main transport mechanism,  the mixing length evolves as $\sqrt{t}$ (the time resolution of the simulations is not enough to show this regime in the vertically stratified media). The onset of the instability makes the mixing length grow faster until a linear growth $(\sim t)$ is observed in the so called non-linear regime \citep{Gopalakrishnan2017}. The variance of the log-permeability field has opposite effects depending on the stratification. The mixing length increases with \slk{} in the vertically stratified cases but decreases in horizontally stratified ones. The case R2, which corresponds to the realistic case of CO$_{2}$ sequestration, takes the highest values compared to the cases R3 and R1. Moreover, it can be observed that the mixing length for vertically stratified cases reaches the bottom of the domain sooner than the horizontally stratified ones in which mixing length growth is slowed down by low permeability layers.

\begin{figure}[htbp]
    \centering \includegraphics[width=0.49\textwidth]{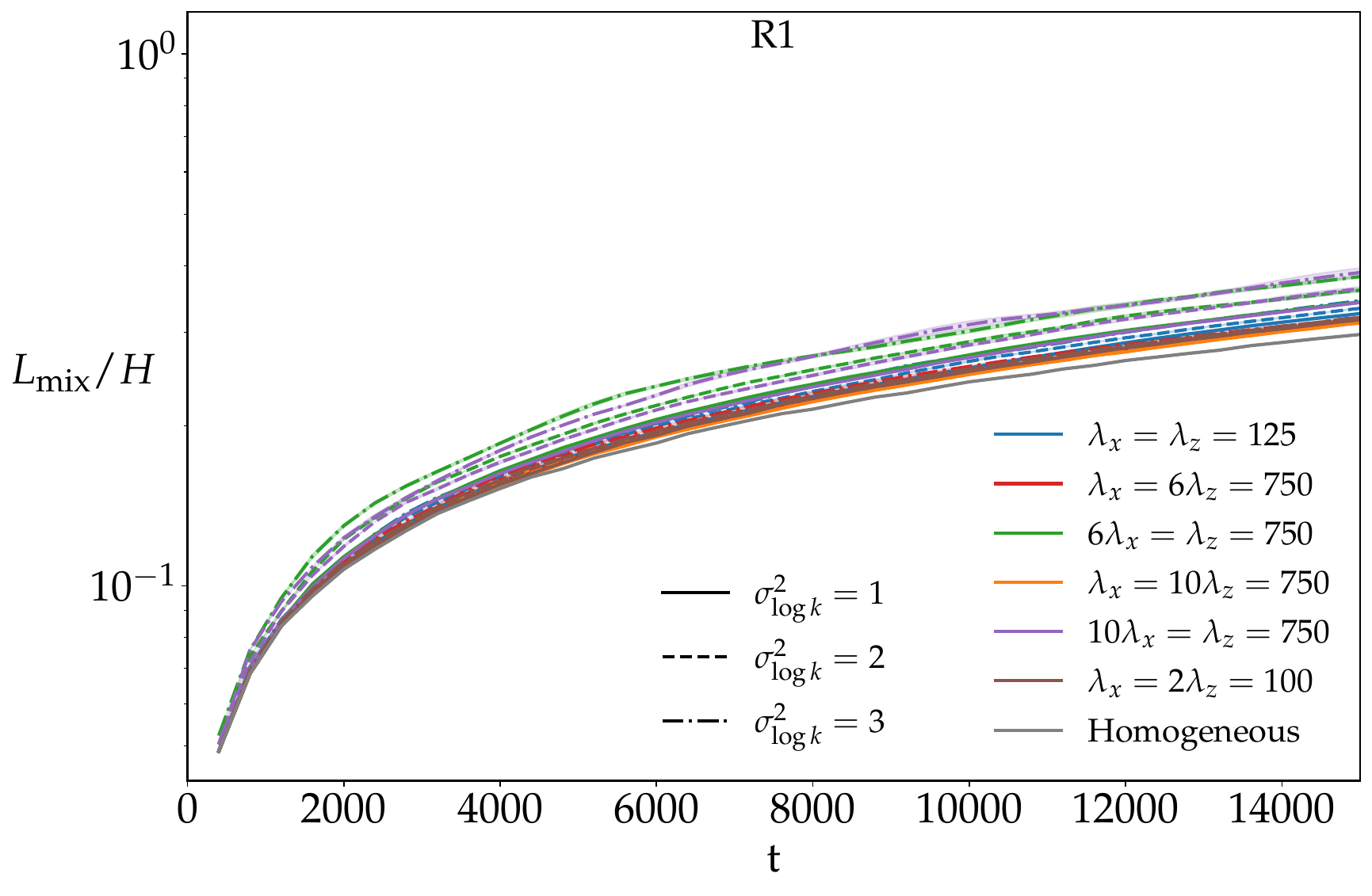}\\
    \includegraphics[width=0.49\textwidth]{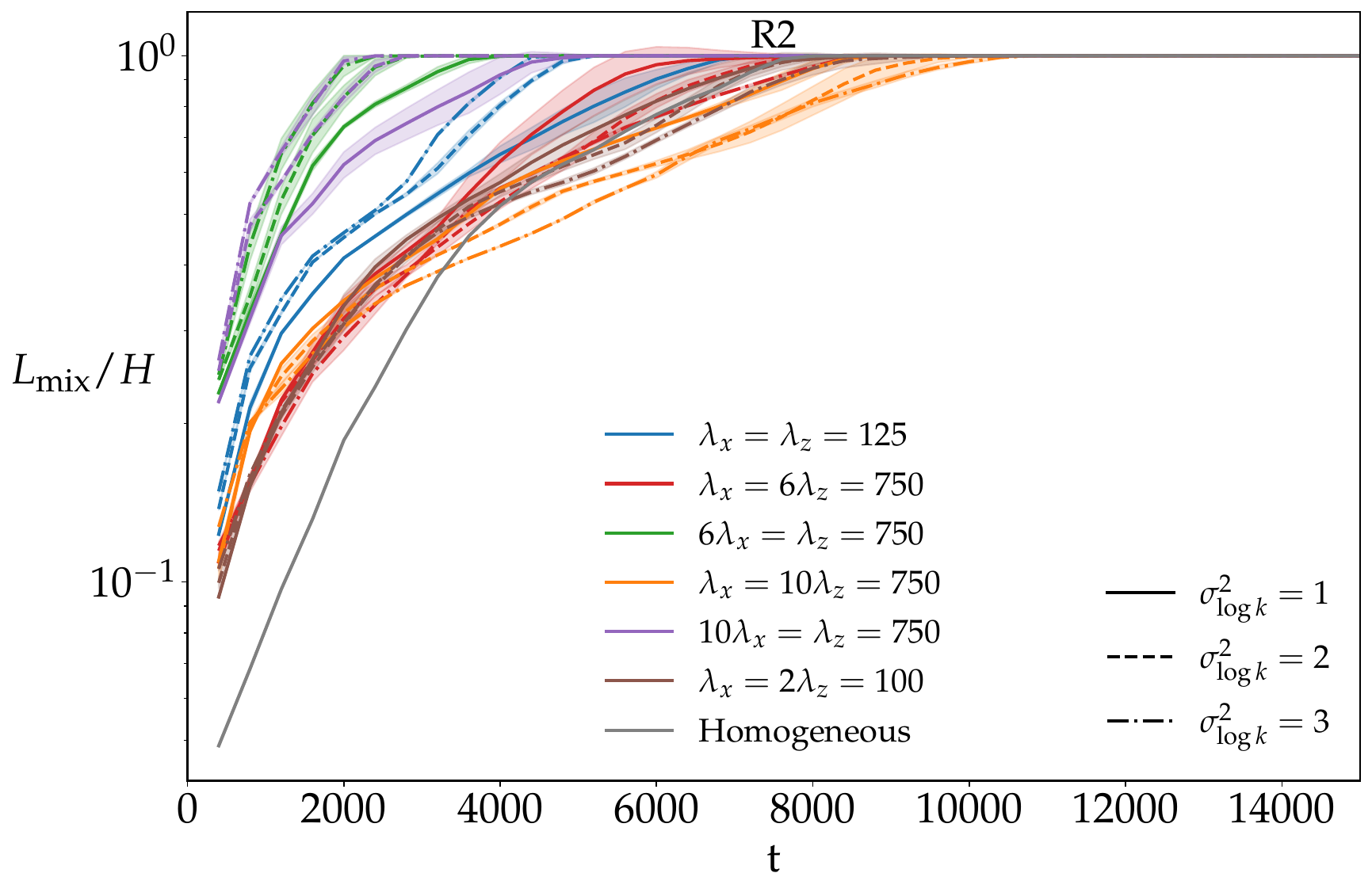}
    \includegraphics[width=0.49\textwidth]{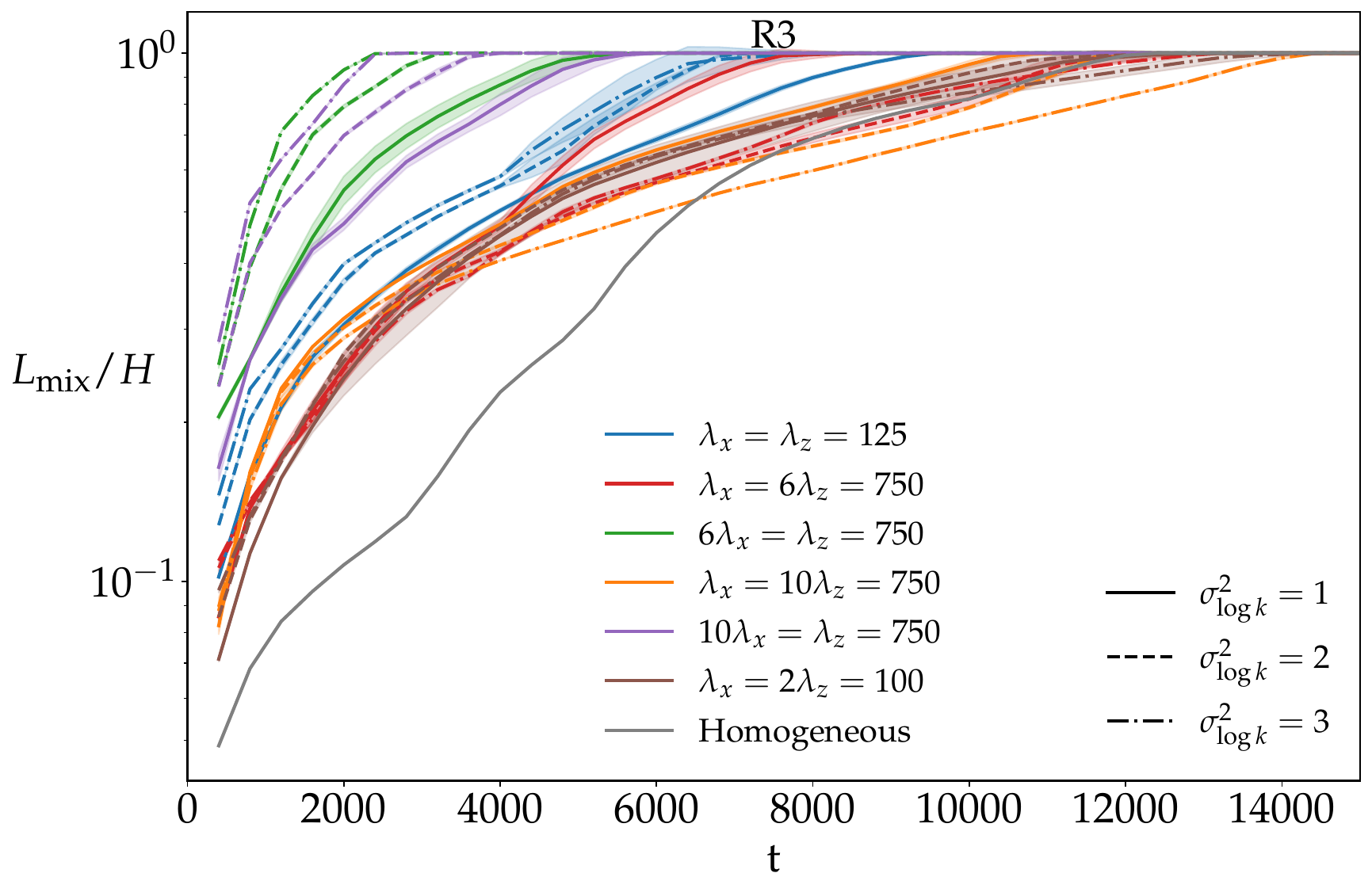}
    \caption{Mixing lengths vs time for the homogeneous and multi-Gaussian heterogeneous cases for all \slk. R1 cases show little sensitivity to changes in \slk, unlike R2 and R3. In R2 and R3 cases, the mixing lengths of the cases with $\lx<\lz$ increase with variance, whereas they decrease when $\lx>\lz$. The shaded area shows the variability between realizations.}
    \label{fig:mixlenlogs}
\end{figure}

Fig.\ref{fig:mixlenlogs} compares the mixing length of the different multi-Gaussian cases. As in the stratified cases, only the R1 case  presents a transition from the diffusive to the non-linear regime. In contrast, cases R2 and R3 do not show evidence of a diffusive regime. This absence is due to a faster transition to the nonlinear regime, which is not captured by the time resolution of the simulations, and to the fact that R2 and R3 cases are more unstable than R1. Similar to the perfectly stratified cases, when $\lx/\lx < 1$, the mixing length grows faster than when $\lx/ \lx > 1$. Although the growth speed decreases sooner for longer \lx because of the barrier effect of low permeability areas. Notably, the isotropic case presents a longer mixing length than the high anisotropy ratio case. All considered multi-Gaussian cases reach the bottom earlier than the homogeneous one except the $\lx = 10\lz=750$ case (purple color lines in Figure \ref{fig:mixlenlogs}), which confirms the controlling role of the vertical correlation length over the evolution of fingers' length. The effect of \slk{} is also similar to that in stratified media, with small anisotropy ratio cases having a mixing length proportional to \slk{} and the opposite behavior is observed for large anisotropy ratio cases.

%
%
\subsection{Reaction and mixing}
In this section, we characterize the reaction zone dynamics by analyzing the reaction rate and front position, the amount of  product and the relation between the reaction rate and the mixing between the reactants.
\subsubsection{Reaction front}
\begin{figure}
    \centering
    \includegraphics[width=\linewidth]{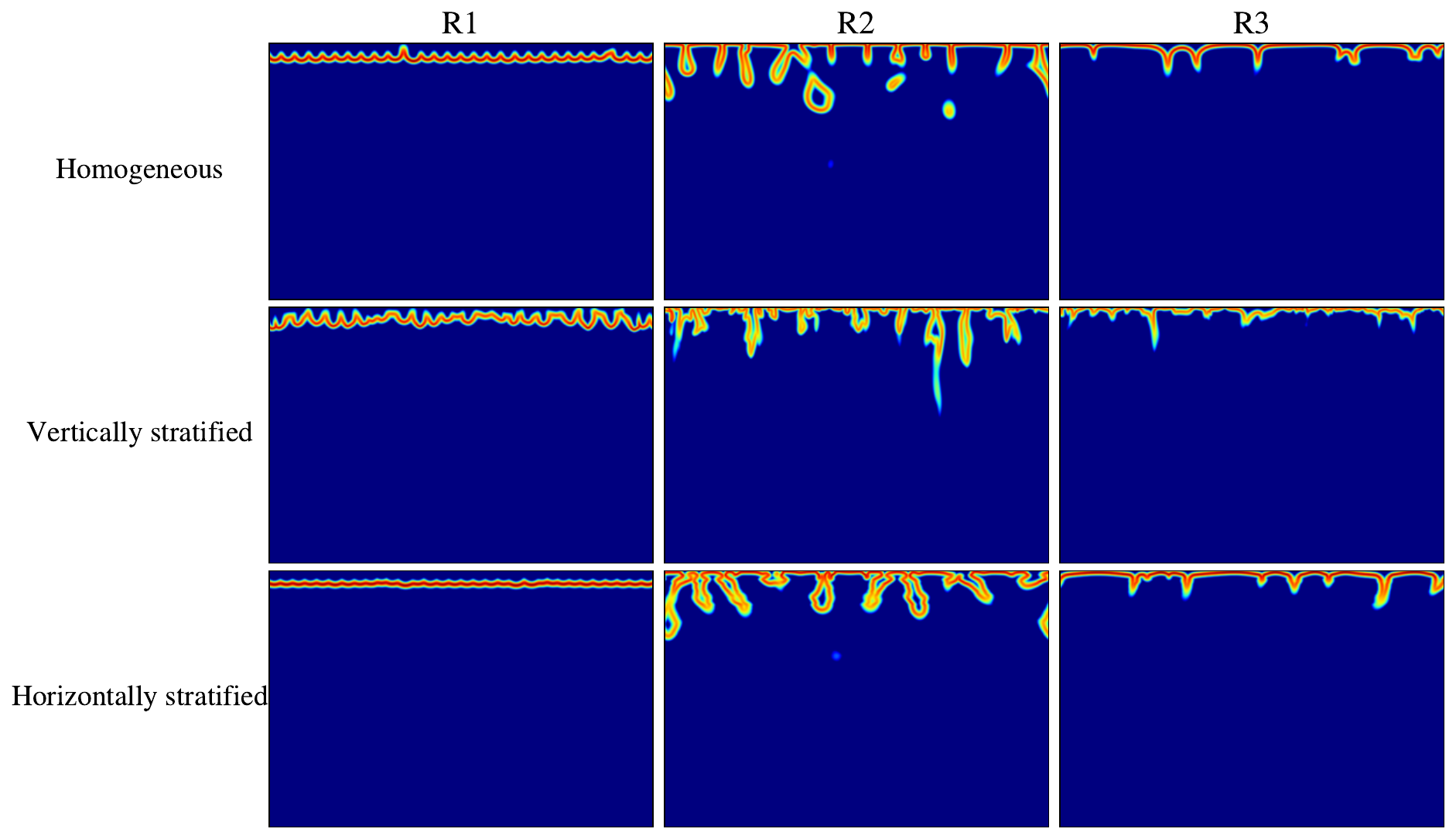}

    \caption{Reaction rate at $t=8000$ for the homogeneous and horizontally stratified media and $t=4000$ for the vertically stratified media. All heterogeneous cases have $\slk=1$. R2 cases front has the fastest speed particularly in the vertically stratified case. The contours of the fingers are the hot spots of the reaction. The front is mostly located at the top of cell for R2 and R3, however is moves faster for R1.}
    \label{fig:reacfront}
\end{figure}

\begin{figure}[ht!]
    \centering
    \includegraphics[width=\linewidth]{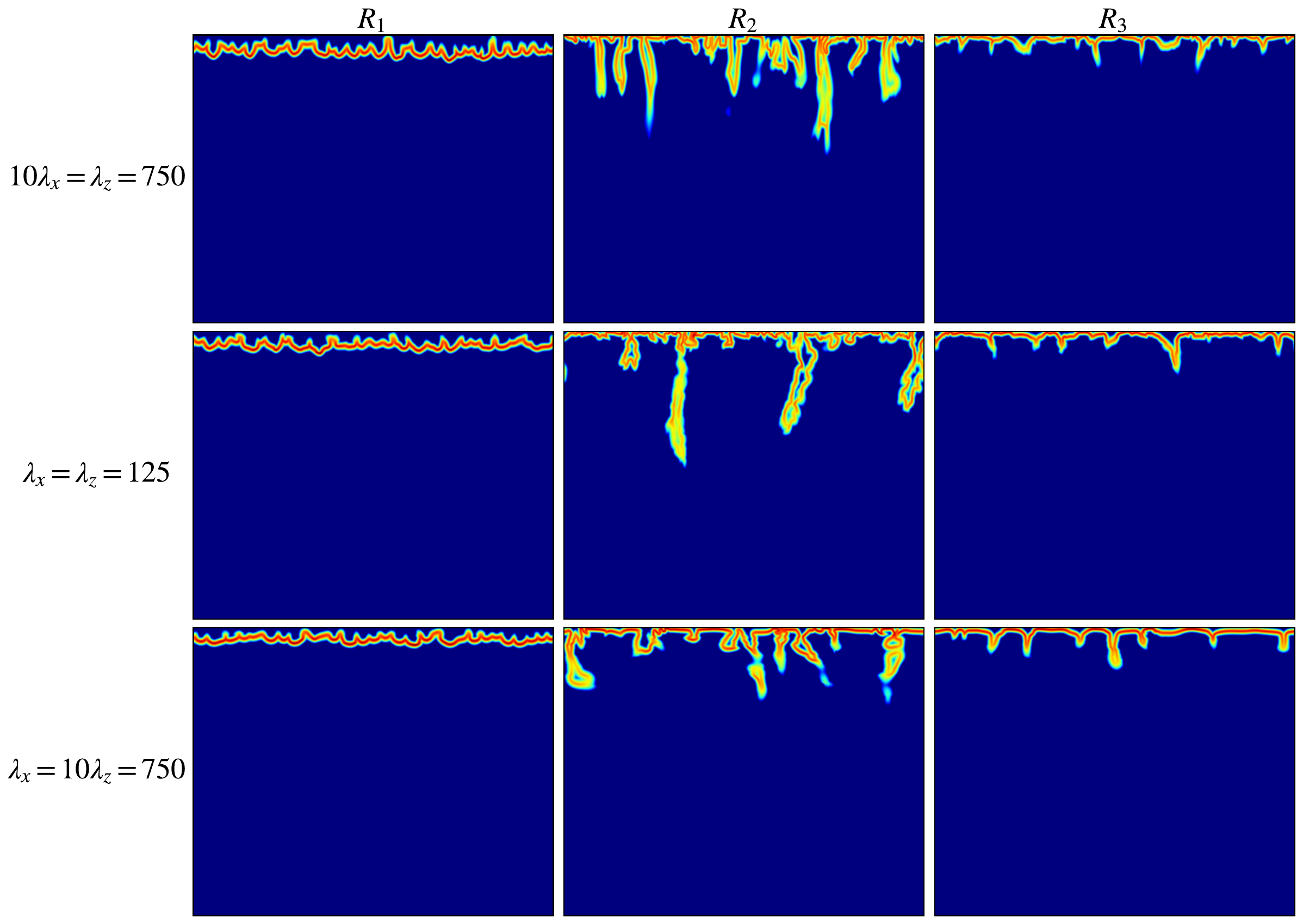}
    \caption{Reaction rate of some of the log-normally distributed cases with different correlation lengths at $t=4000$ with $\slk=1$. Cases with $\lx<\lz$ have thinner longer fingers, whereas cases with $\lx>\lz$ have thicker, laterally extended fingers. The hot spots of the reaction are mainly located at the top.}
    \label{fig:reacfrontR2log}
\end{figure}

We define the reaction rate as
\begin{align}
  \label{eq:reaction_rate}
  r(x,z,t)  = \cA(x,z,t) \cB(x,z,t)
\end{align}
and the reaction front as the region where $r(x,z,t)>0$ (see figures \ref{fig:reacfront} and \ref{fig:reacfrontR2log}). We characterize the reaction front by the position of its center of mass $z_{f}$ and width $w_{f}$ defined using the first and second moments as \citep{Dentz2022}

\begin{align}
  \label{eq:center_mass}
  z_{f} (t)= \overline{\frac{m^{1}(x,t)}{m^{0}(x,t)} }
\end{align}
\begin{align}
  \label{eq:width}
  w_{f}(t) = \overline{\frac{m^{2}(x,t)}{m^{0}(x,t)} - \left(\frac{m^{1}(x,t)}{m^{0}(x,t)}\right)^{2}},
\end{align}
where
\begin{align}
  \label{eq:moment}
  m^{j}(x, t) = \int_{0}^{H}  z^{j} r(x,z,t) dz
\end{align}
is the moment of order $j$ and the bar $(\overline{\cdot})$ indicates averaging over the $x$ direction.

\begin{figure}
  \centering 
 \textbf{Vertically stratified} \\[0.5ex]
\includegraphics[width=\linewidth]{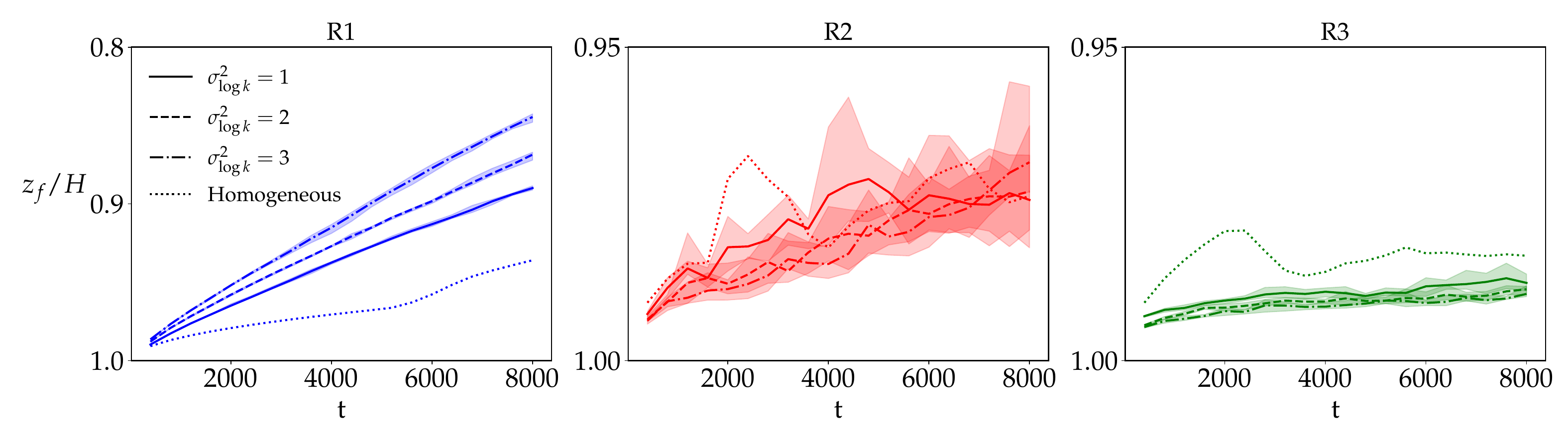}

\vspace{-1.5ex}  

\textbf{Horizontally stratified} \\[0.5ex]
\includegraphics[width=\linewidth]{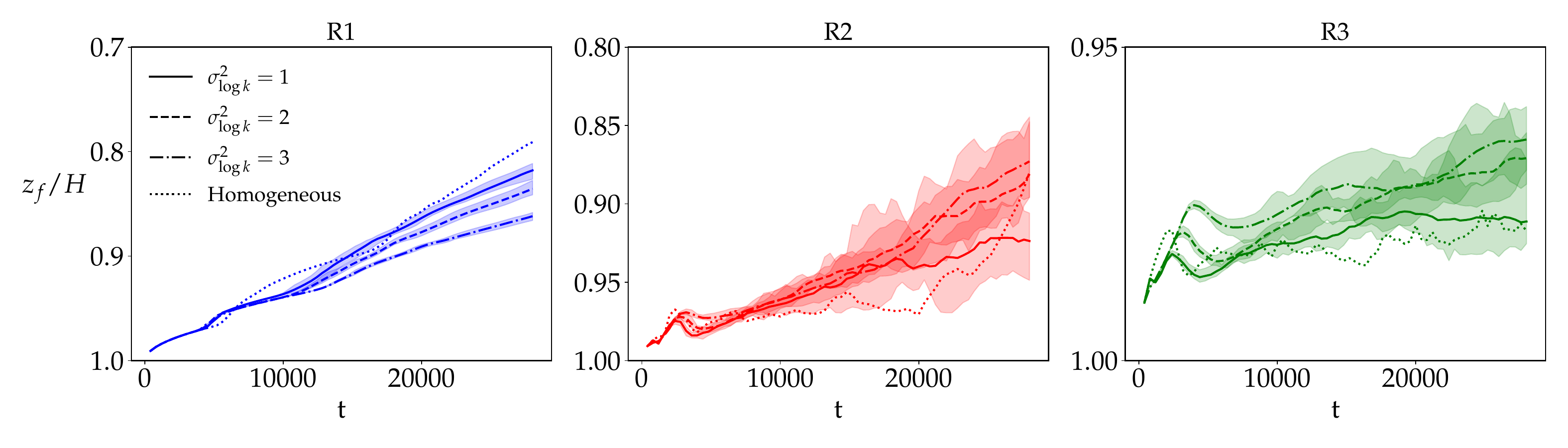}

  \caption{Reaction front position versus time for the homogeneous and all stratified cases. The R1 front advances more rapidly than in the R2 and R3 cases and gets steeper with a larger variance. However, in R2 and R3, fronts are compact and chemically active regions are located at the top of the cell (see figure \ref{fig:reacfront}). Vertically stratified reaction fronts propagate faster than their homogeneous counterparts for R1 but slower for R2 and R3; the opposite trend is observed for the horizontally stratified cases. Besides, increasing \slk{} tends to make the front position narrower for vertically stratified cases R2 and R3 (unlike R1) whereas the opposite trend is observed for the horizontally stratified ones. The shaded area shows the variability between realizations.}
    \label{fig:frontposits}
\end{figure}

\begin{figure}
    \centering
    \includegraphics[width=\linewidth]{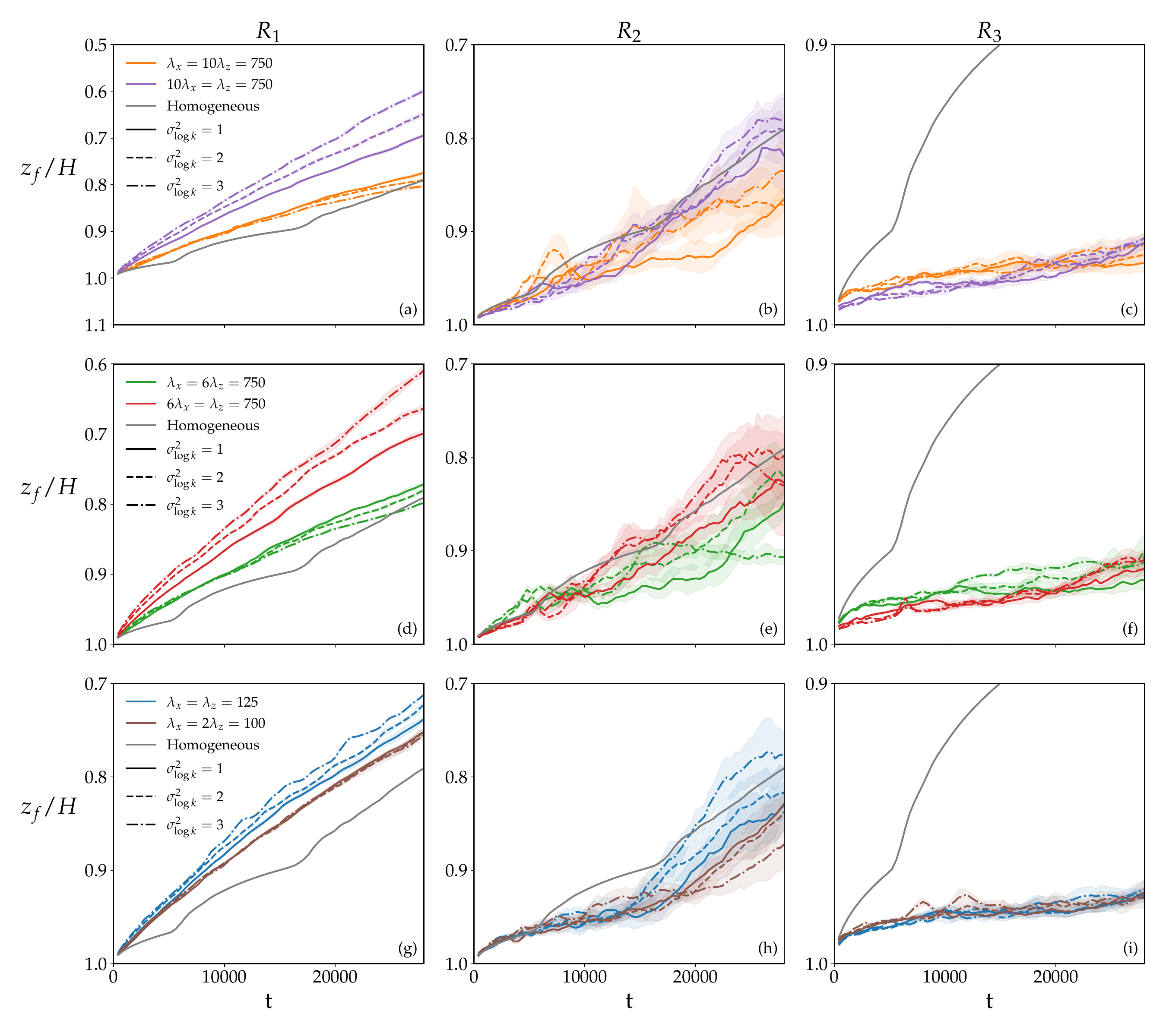}
    
    \caption{Reaction front position versus time for the homogeneous and multi-Gaussian cases. The front position follows the same trend observed for previous parameter values. Cases with $\lx<\lz$ have a faster front proportionally to the variance whereas the opposite trend is observed for cases where $\lx>\lz$. Furthermore, the isotropic case behaves in the same way as cases $\lx<\lz$. In contrast to R1, the reaction fronts in R2 and R3 advance more slowly than their homogeneous counterparts. The shaded area shows the variability between realizations.}
    \label{fig:frontpositlogs}
\end{figure}

Fig.\ref{fig:frontposits} shows the temporal evolution of the position of the reaction front for the homogeneous and stratified cases. Unlike the mixing length, fronts corresponding to vertically stratified media do not advance faster that the ones in horizontally stratified media in all cases. Only for the case R1 does the front advance faster. In the R2 and R3 cases, the homogeneous and horizontally stratified media behave in a similar manner with a fast advance followed by a rebound after the onset of convection. The oscillations in the position of the front are larger for the horizontally stratified case because of the effect of the alternate high and low permeability layers. R1 cases present less fluctuations due to the fact that the reaction product does not contribute to density, and the front stops advancing at some point when the minimum in density  stabilizes the system. Vertically stratified reaction fronts propagate faster than their homogeneous counterparts for R1 but slower for R2 and R3; the opposite trend is observed for the horizontally stratified cases.  In the multi-Gaussian cases (Figure~\ref{fig:frontpositlogs}), the front advances deeper than in the homogeneous case. Moreover, scenarios with $\lx/\lz < 1$ exhibit greater spatial extension than those with an anisotropy ratio greater than 1, with the isotropic case lying between them. Additionally, cases R2 and R3 show more pronounced fluctuations compared to R1, due to their increased instability.

The effect of \slk{} on the front position and width varies depending on the stratification. Vertically stratified reaction fronts propagate faster than their homogeneous counterparts for R1 but slower for R2 and R3 whereas the opposite trend is observed for the horizontally stratified cases. The same behavior is observed for the multi-Gaussian cases with the largest and smallest anisotropy ratio.


\begin{figure}
    \centering
     \textbf{Vertically stratified} \\[0.5ex]
\includegraphics[width=\linewidth]{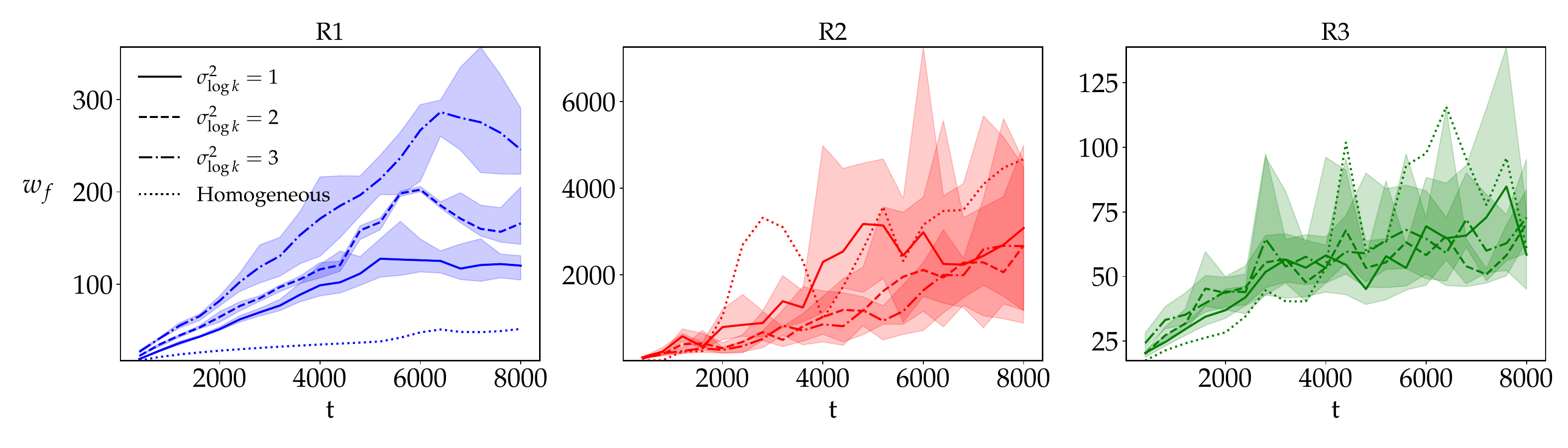}

\vspace{-1.5ex}  

\textbf{Horizontally stratified} \\[0.5ex]
\includegraphics[width=\linewidth]{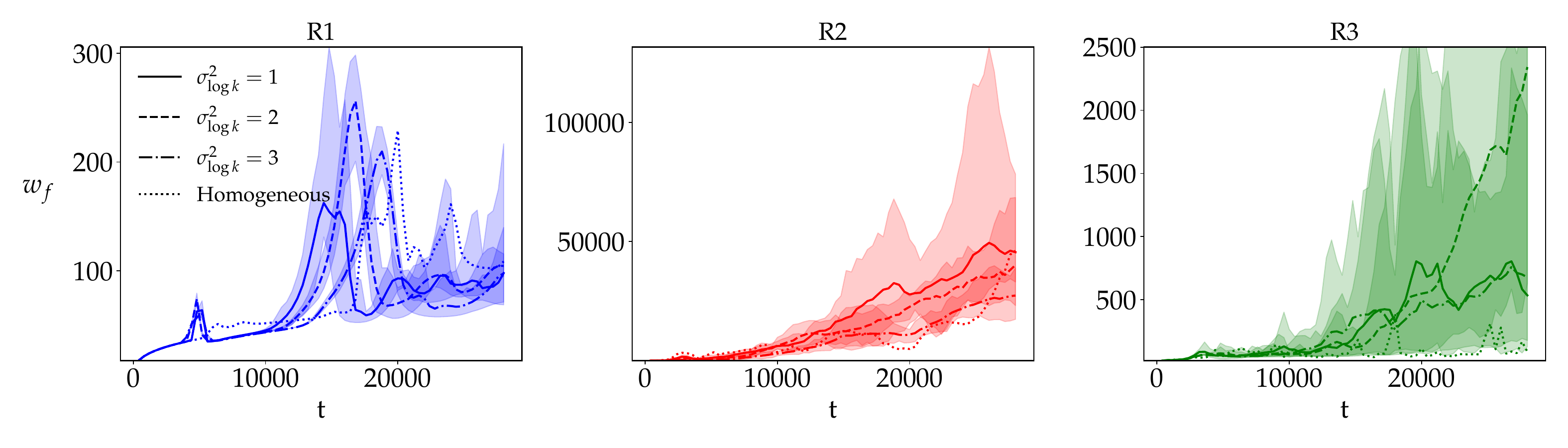}

    \caption{Reaction front width versus time for the homogeneous and all stratified cases. Horizontally stratified cases possess wider R2 reaction fronts than vertically stratified ones which in turn are narrower than the homogeneous R2 front. Furthermore, the front width decreases with \slk{} for both cases. Nonetheless, no clear tendency is observed for R1 and R3. The shaded area shows the variability between realizations.} 
    \label{fig:frontwmeans}
\end{figure}

\begin{figure}[htbp]
    \centering
    \includegraphics[width=\linewidth]{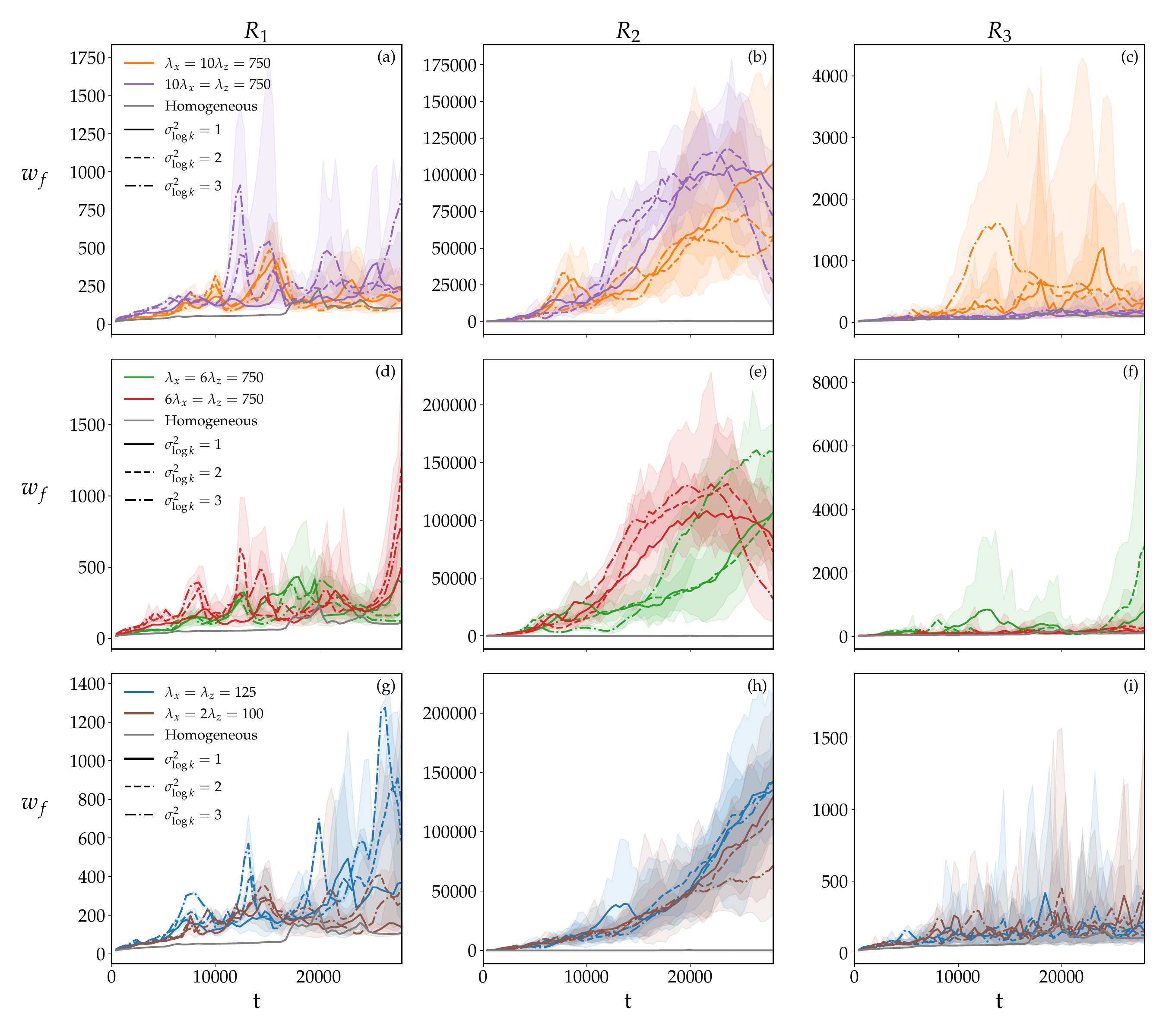}
    
    \caption{Reaction front width versus time for the homogeneous and multi-Gaussian cases. As previously noted for the other considered parameters, the front width also increases with \slk{} for cases with $\lx<\lz$ and decreases proportionally to \slk{} for $\lx>\lz$. Furthermore, the isotropic case behaves in the same way as cases $\lx<\lz$. R2 cases present the widest fronts. The shaded area shows the variability between realizations. Here, the shaded regions represent the range between the minimum and maximum values across realizations.}
    \label{fig:frontwmeanlogs}
\end{figure}

As for the front width (figures \ref{fig:frontwmeans} and \ref{fig:frontwmeanlogs}), R2 cases have the widest front for all cases, being wider in the horizontally stratified ones than in vertically stratified ones (which are even less wide than the homogeneous R2 front), because of the lateral spreading of the fingers. The front width decreases with \slk{} for both R2 stratified cases. However, no clear tendency is observed for R1 and R3 in the horizontally stratified cases. In multi-Gaussian fields (see figures \ref{fig:frontpositlogs} and \ref{fig:frontwmeanlogs}), the relationship between \slk{} and front dynamics also reverses as the anisotropy ratio changes. In cases with $\lx/\lz <1$ (purple and red in panels (a) to (f) in figure \ref{fig:frontwmeanlogs} and \ref{fig:frontpositlogs}), the front width and position increase proportionally to the variance unlike when $\lx/\lz>1$ (orange and green in panels (a) to (f) in figures \ref{fig:frontwmeanlogs} and \ref{fig:frontpositlogs}), where the opposite trend is observed.

%
%
\subsubsection{Reaction rate and mass of product}
To further understand the effect of heterogeneity on the reaction dynamics, we examine the evolution of the total reaction rate $\langle r(t)\rangle$ and the total amount of the reaction product $m_{\C}(t)$ defined as
\begin{align}
  \label{eq:total_reaction_rate}
\langle r(t)\rangle = \int_{\Omega} r(x,z, t)  d\Omega
\end{align}
\begin{align}
  \label{eq:mc}
m_{\C}(t) = \int_{\Omega} \cC(x,z,t)  d\Omega.  
\end{align}
\begin{figure}
    \centering
    \includegraphics[width=0.49\linewidth]{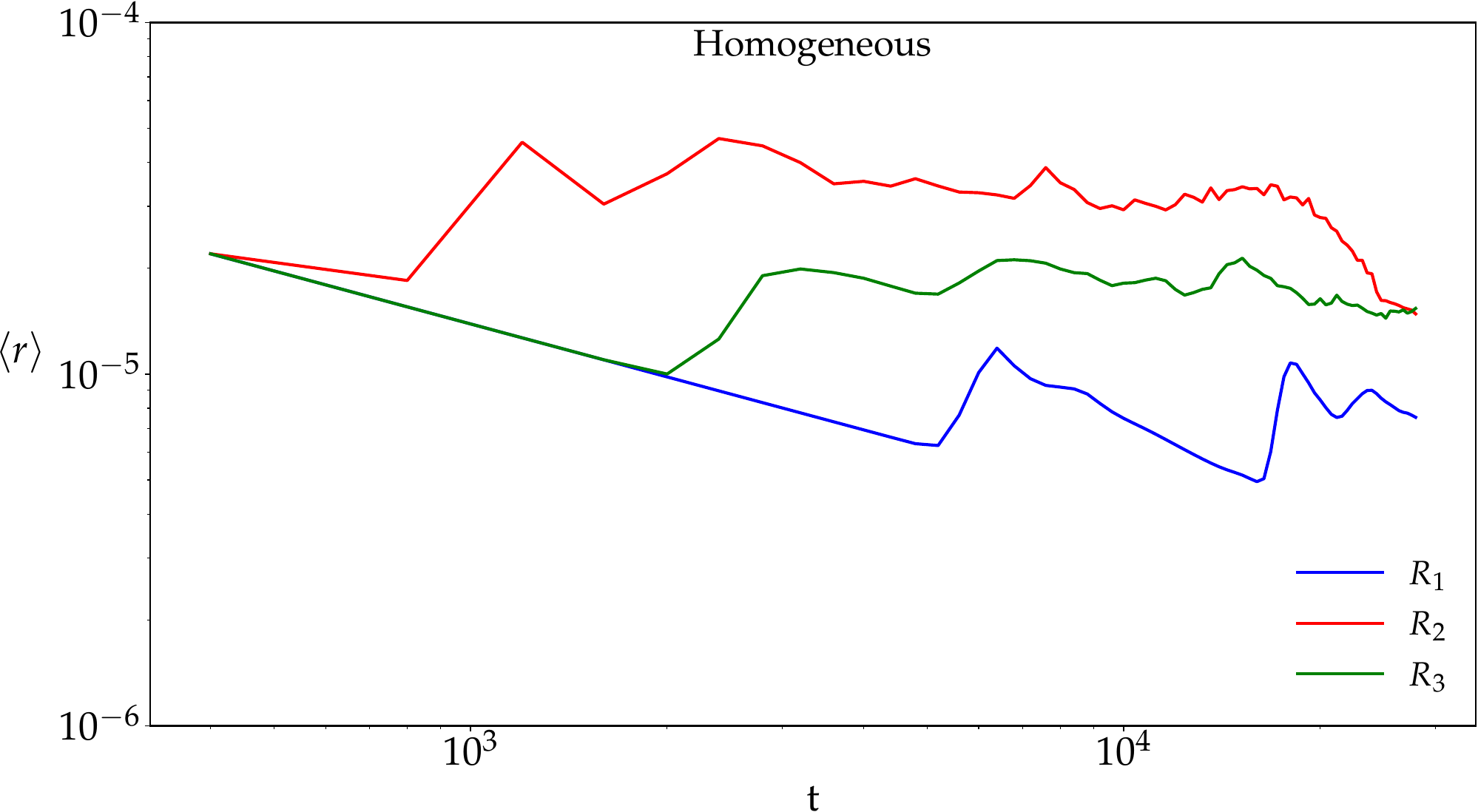}\\
    \includegraphics[width=0.45\linewidth]{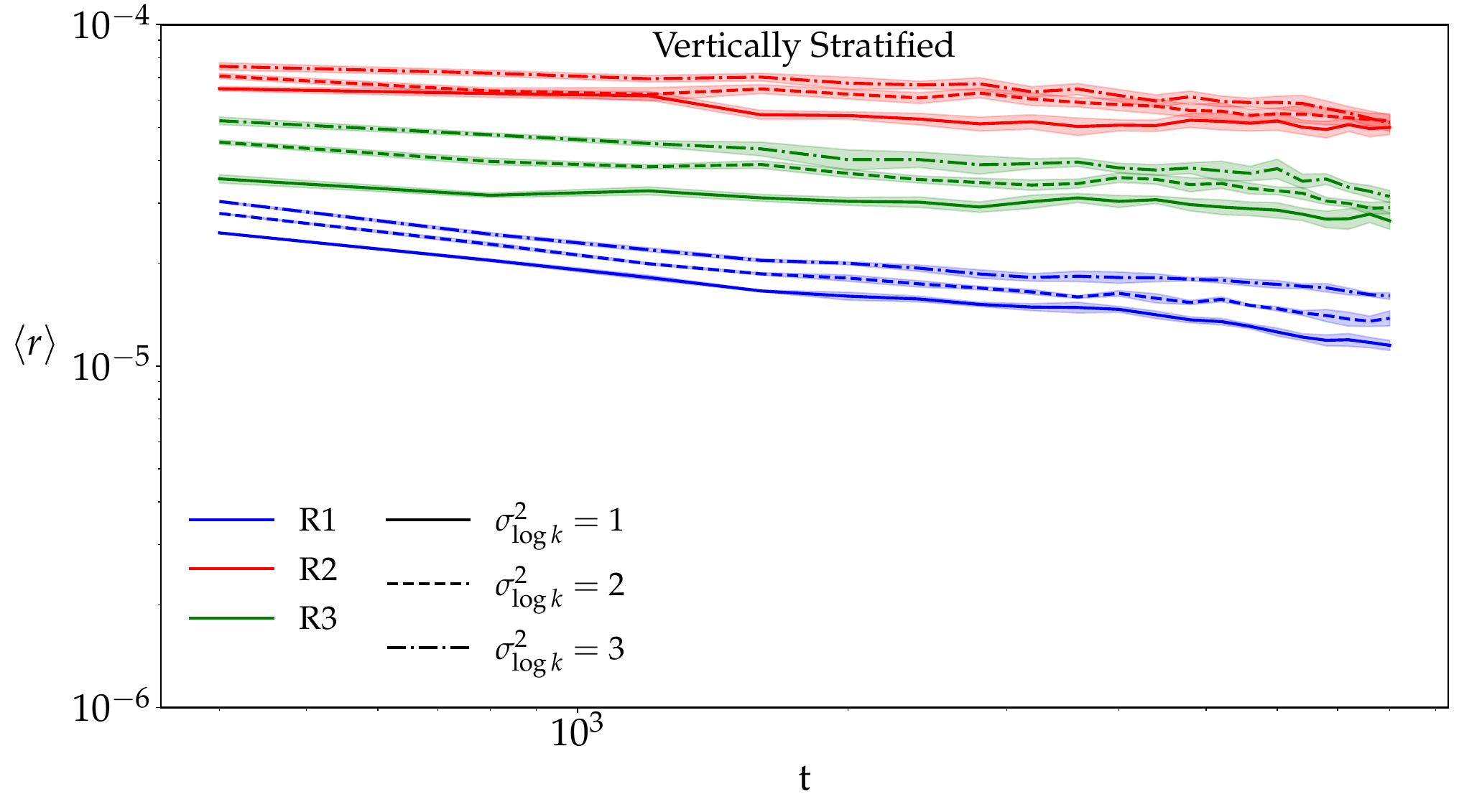}
    \includegraphics[width=0.45\linewidth]{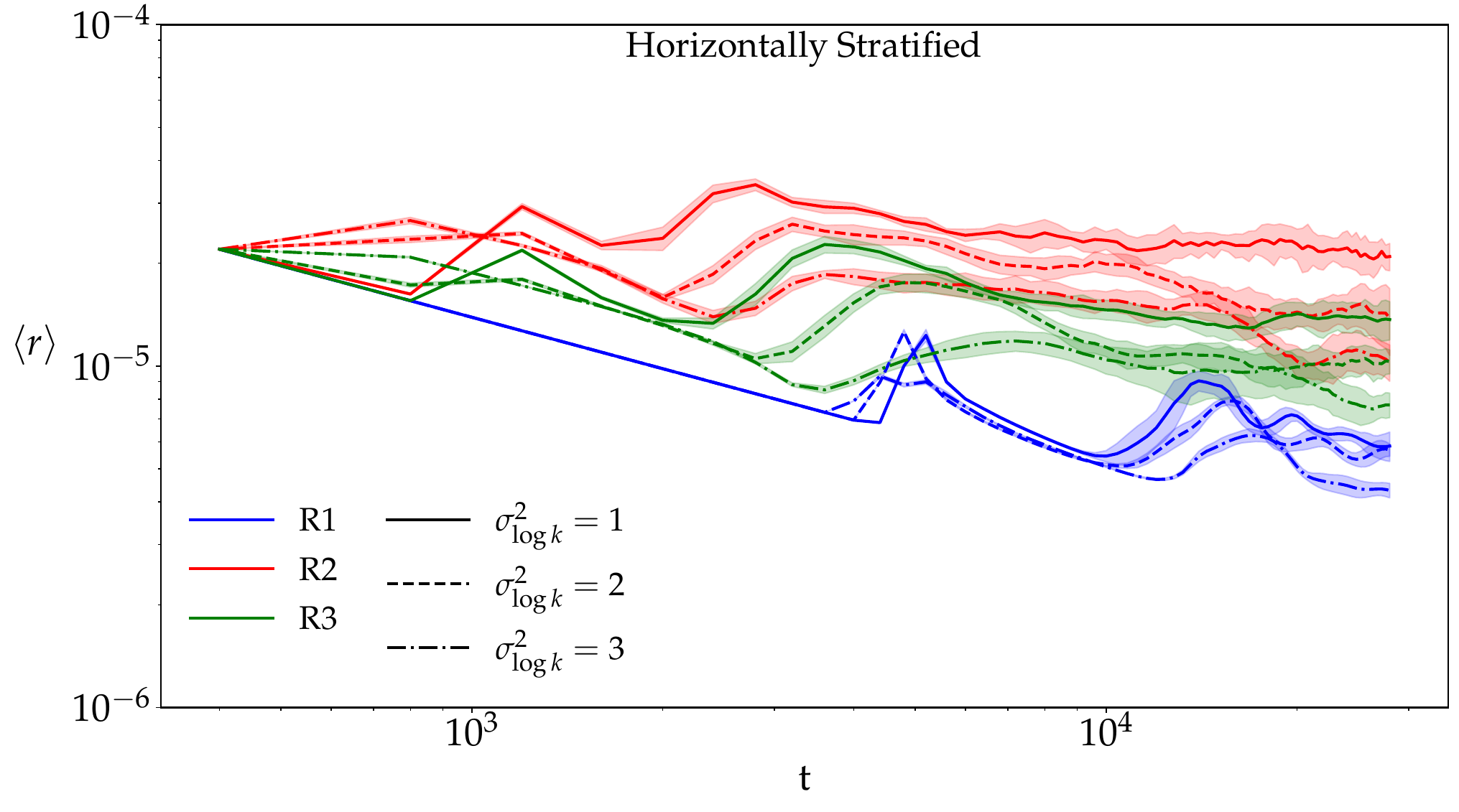}
    \caption{Reaction rate versus time for the homogeneous and all stratified cases. Vertically stratified reaction rates increase with variance whereas the horizontally stratified ones present the opposite behavior. R2 cases are the most chemically active cases followed by R3 then R1 The shaded area shows the variability between realizations.}
    \label{fig:rvsts}
\end{figure}

\begin{figure}
  \centering
  \includegraphics[width=\linewidth]{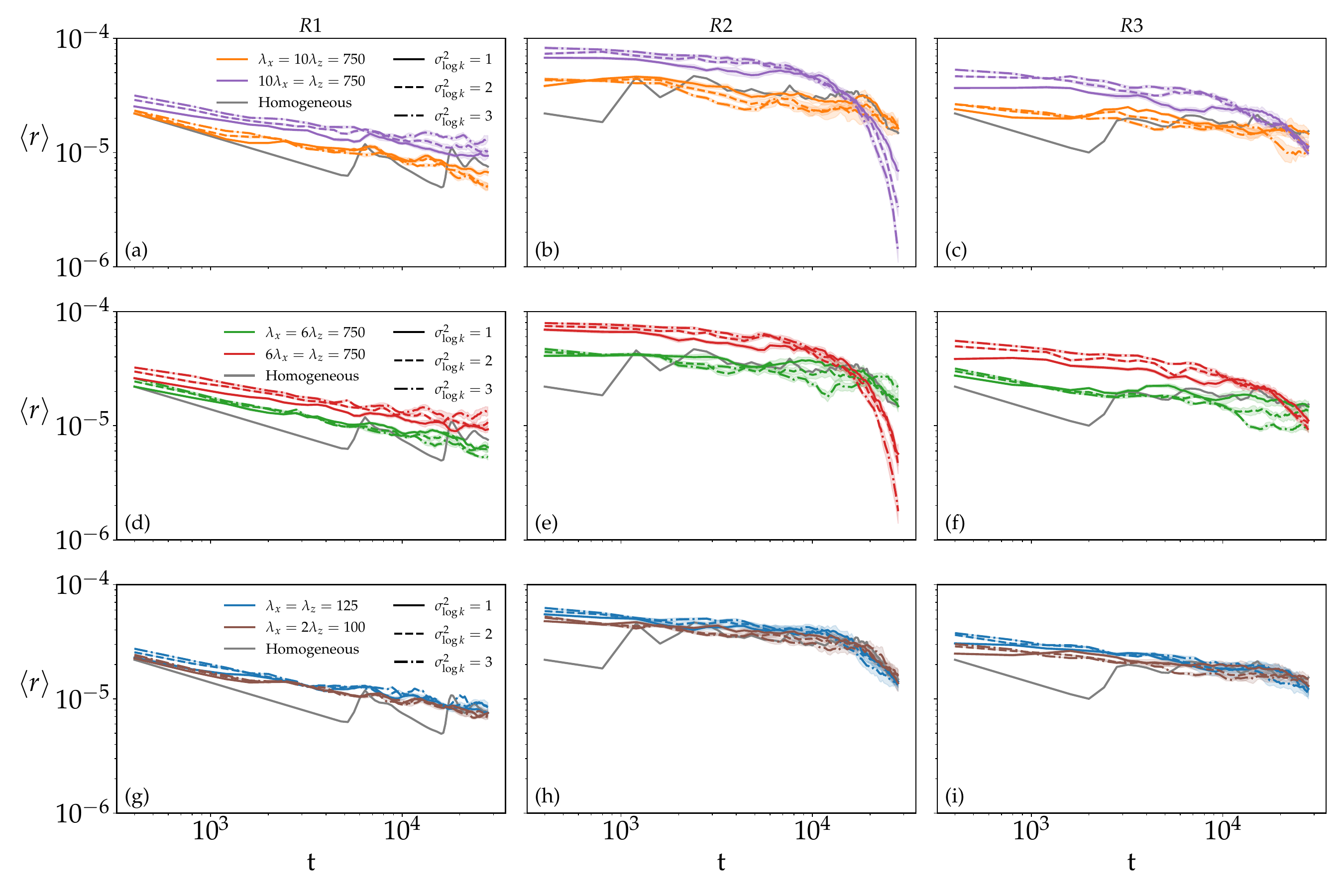}   
    \caption{Reaction rate versus time for the homogeneous and multi-Gaussian heterogeneous cases. The reaction rate increases with \slk{} for cases with $\lx<\lz$ and decreases proportionally to \slk{} for $\lx>\lz$. Furthermore, the isotropic case exhibits similar behavior to cases with $\lx<\lz$ and presents values that are close the case $\lx=2\lz$=100. Additionally, for R2 and R3, the trends are reversed once the simulation reaches the time where the fingers attain the bottom boundary; beyond this point, the results no longer represent meaningful physical behavior. The shaded area shows the variability between realizations.}
    \label{fig:rvstlog}
\end{figure}

The reaction rate (figures \ref{fig:rvsts} and \ref{fig:rvstlog}) reaches larger values in the vertically stratified cases than in the horizontally stratified and homogeneous ones, R2 being the most reactive one. Contrary to what is observed for the mixing length and front dynamics, reaction rate is proportional to \slk. For the multi-Gaussian cases, $\langle r(t)\rangle$ depends on \lx and the anisotropy ratio of the permeability field. When \lx is large, the barrier effect of the low permeability areas reduces the reaction rate with only a minor influence from the anisotropy ratio. When \lx is small, the reaction rate increases, but also presents an earlier shutdown regime as the fingers reach the  bottom of the domain quickly. The dependency with \slk{} is not as clear as in the stratified cases as the behavior of $\langle r(t)\rangle$ presents more oscillations. However, at early times, it follows the same scaling as the stratified media, especially for the cases with the highest and lowest anisotropy ratio. 

\begin{figure}
    \centering
    \includegraphics[width=0.49\linewidth]{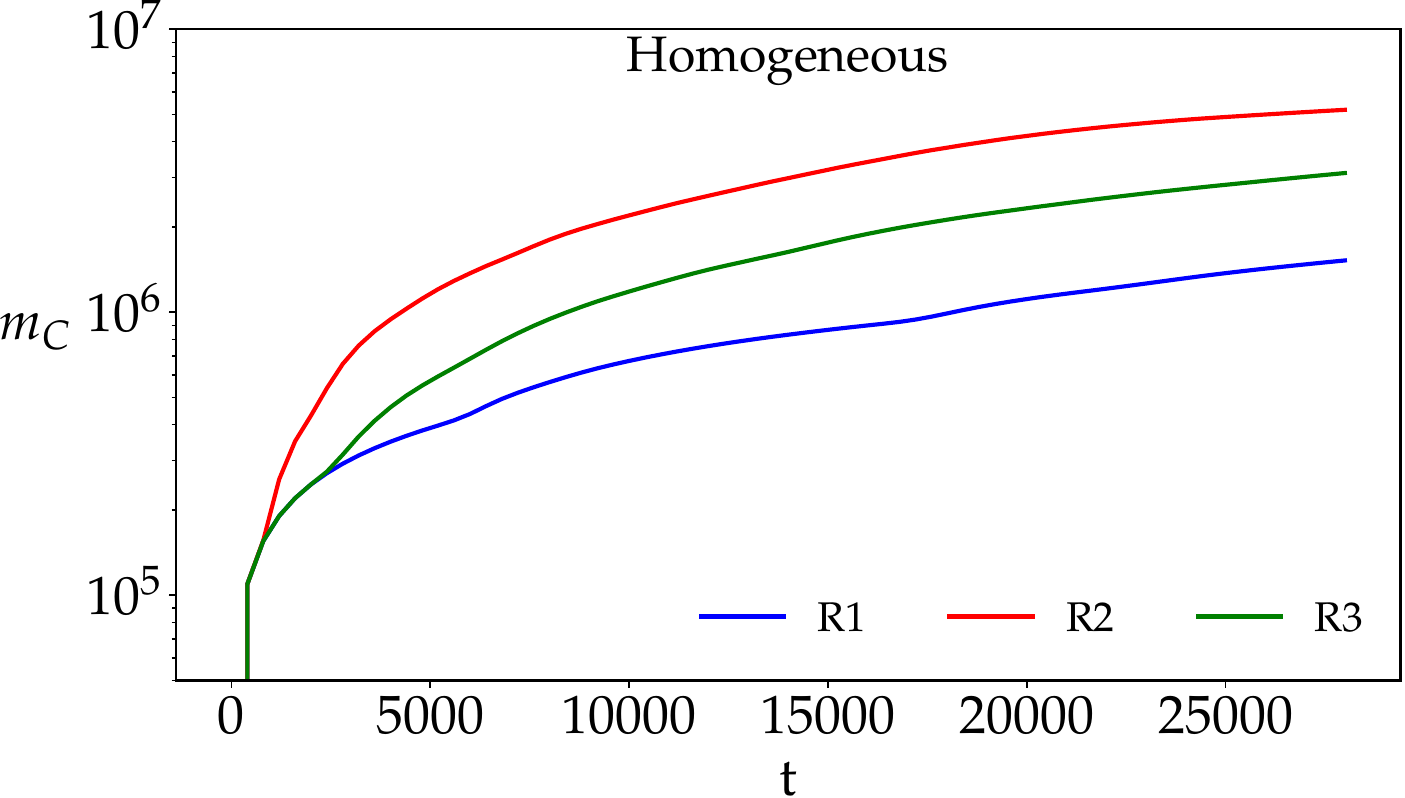}\\
    \includegraphics[width=0.49\linewidth]{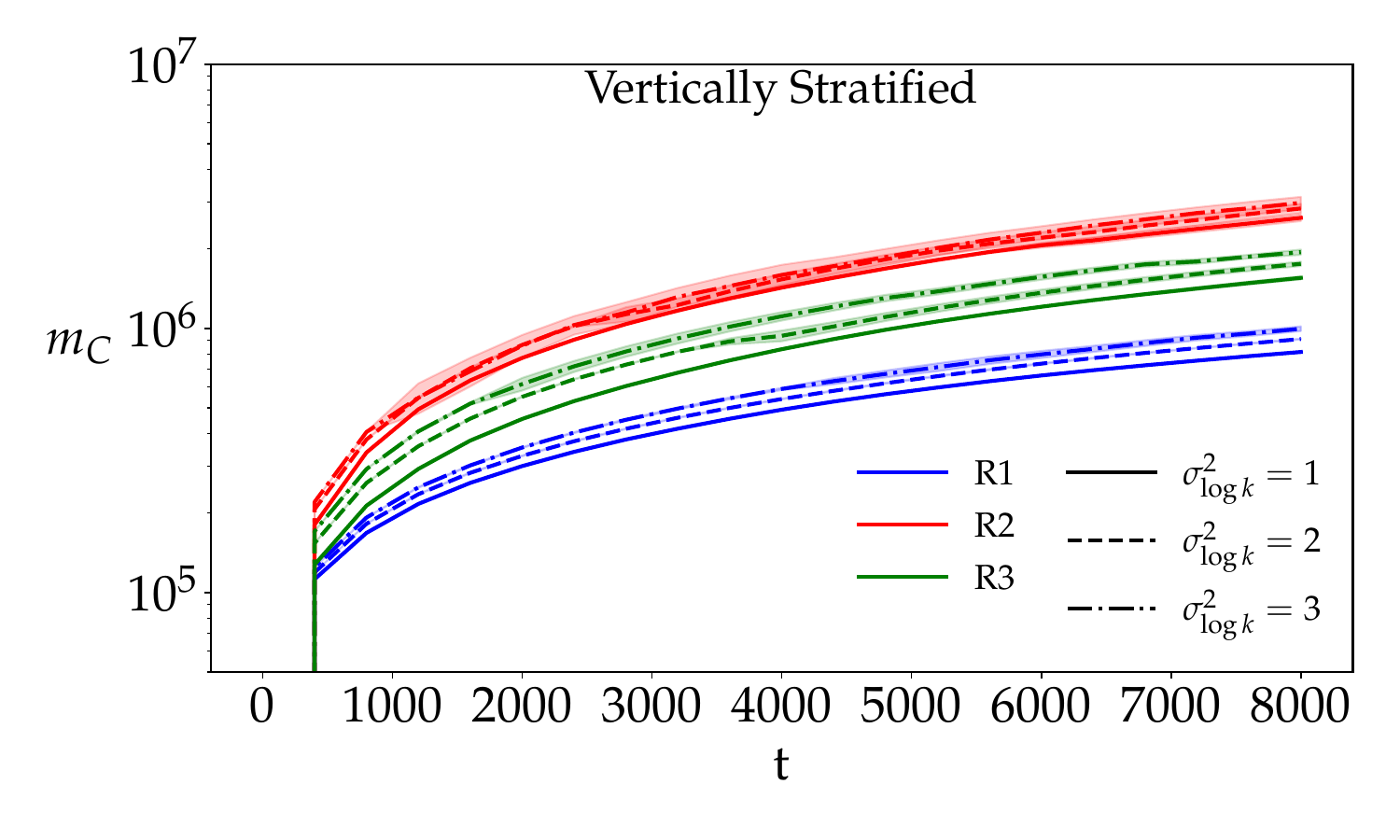}
    \includegraphics[width=0.49\linewidth]{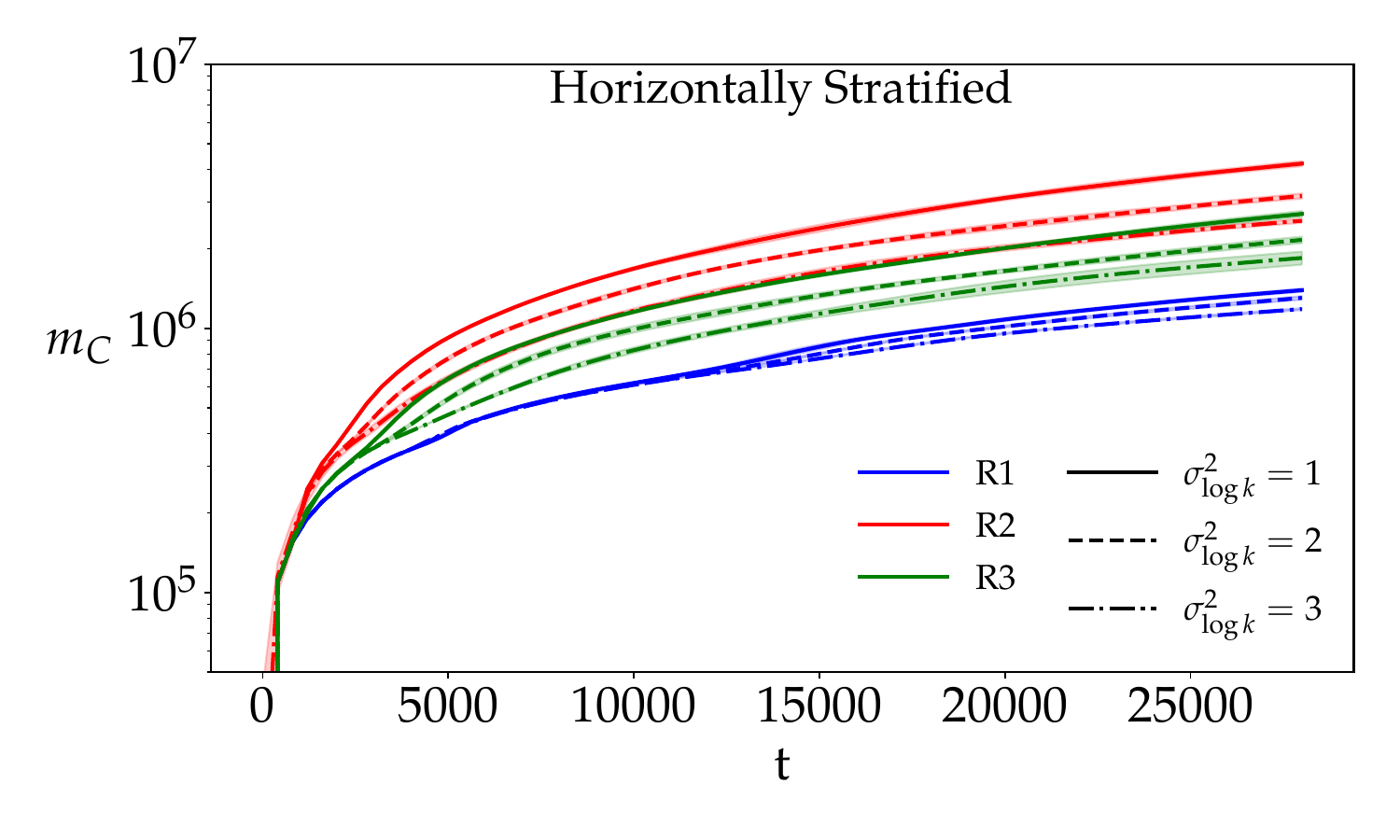}
    \caption{Amount of product for the homogeneous and stratified cases. The mass of product behaves as the mixing length (Figure \ref{fig:mixlencompstrats}). $m_{\C}$ is proportional to $\slk$ in the vertically stratified cases and inversely proportional to $\slk$ in the horizontally stratified ones. The shaded area shows the variability between realizations.}
    \label{fig:mcstratcomp}
\end{figure}

\begin{figure}
    \centering
    \includegraphics[width=0.3\linewidth]{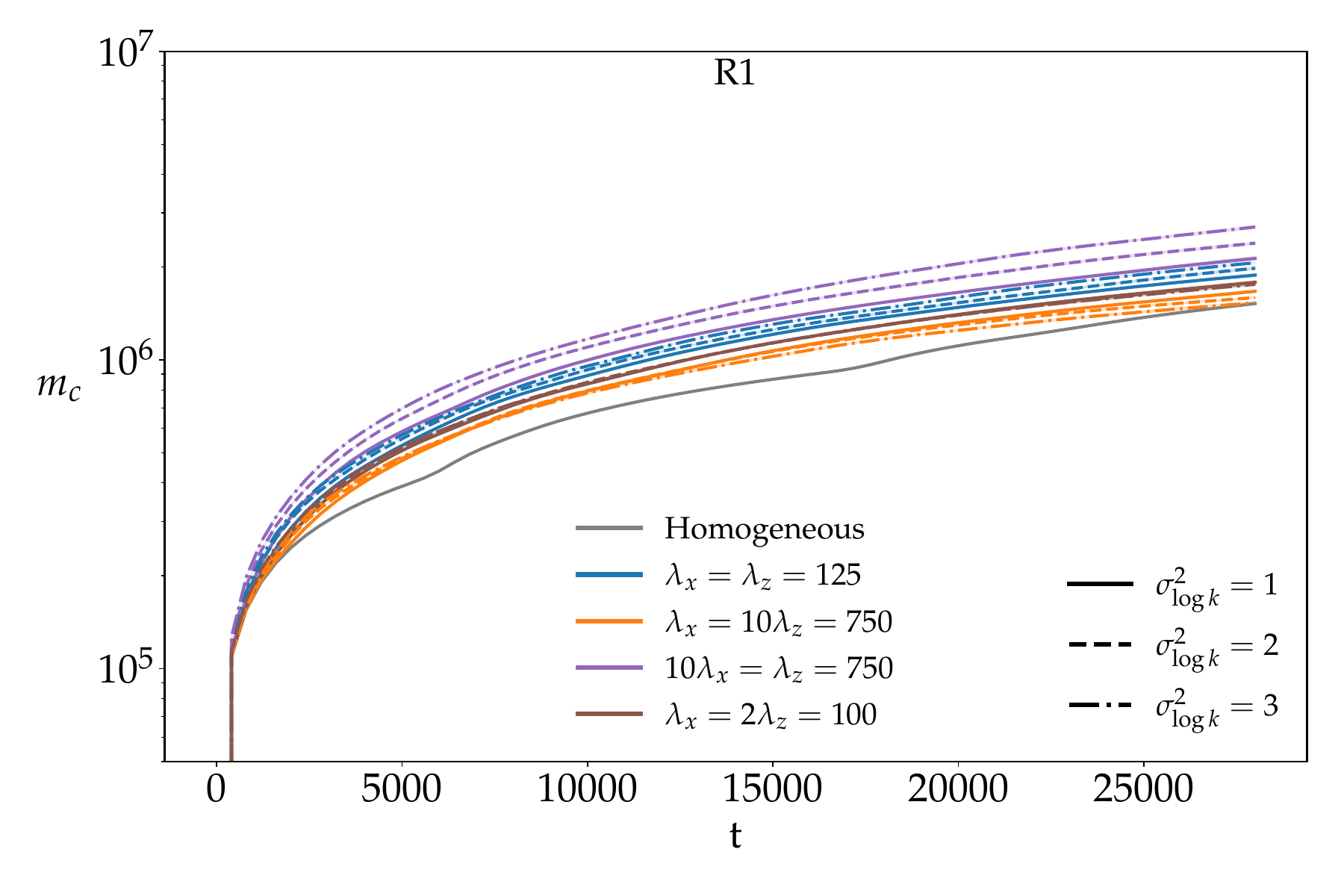}
    \hspace{0.015\linewidth}
    \includegraphics[width=0.3\linewidth]{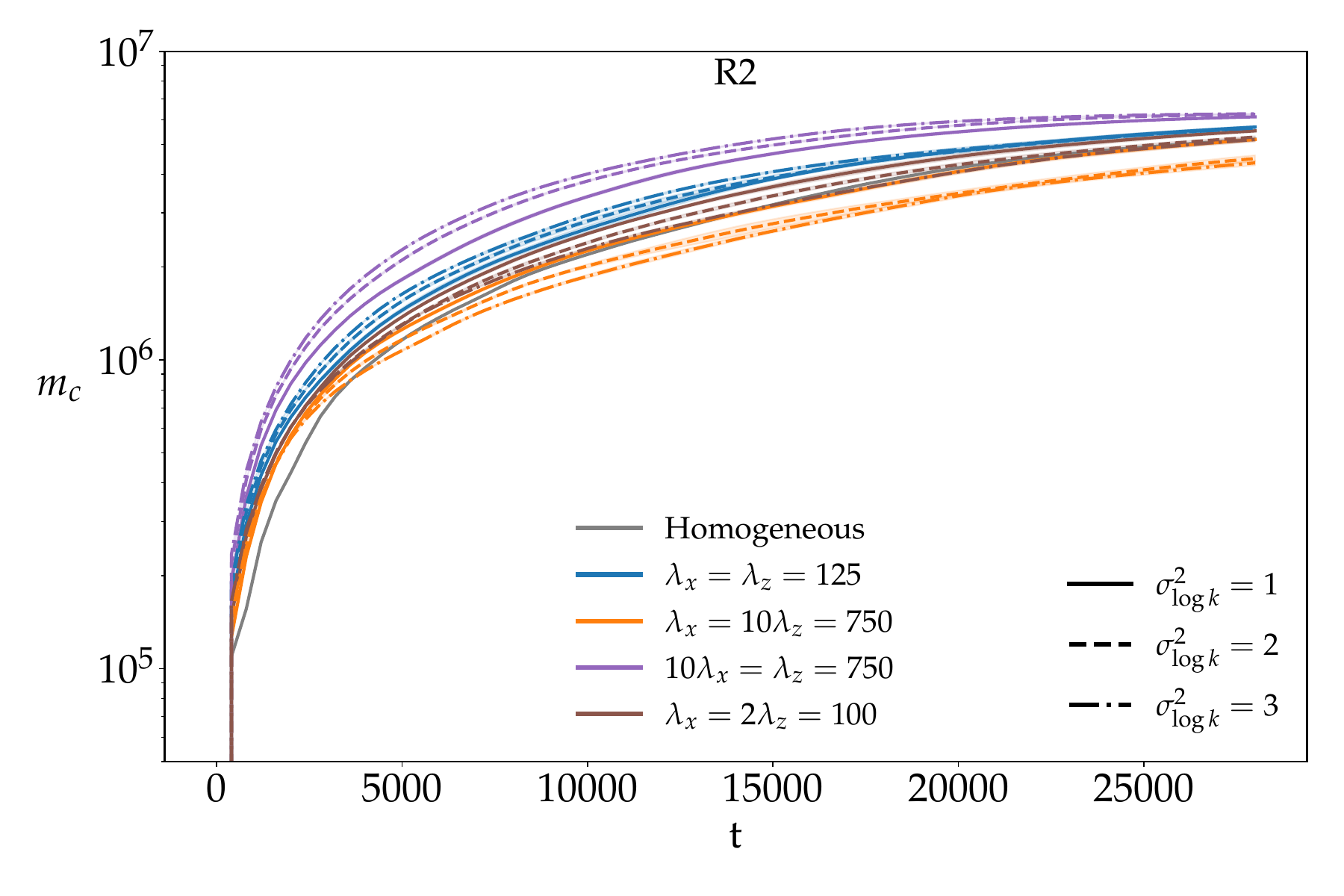}
    \hspace{0.015\linewidth}
    \includegraphics[width=0.3\linewidth]{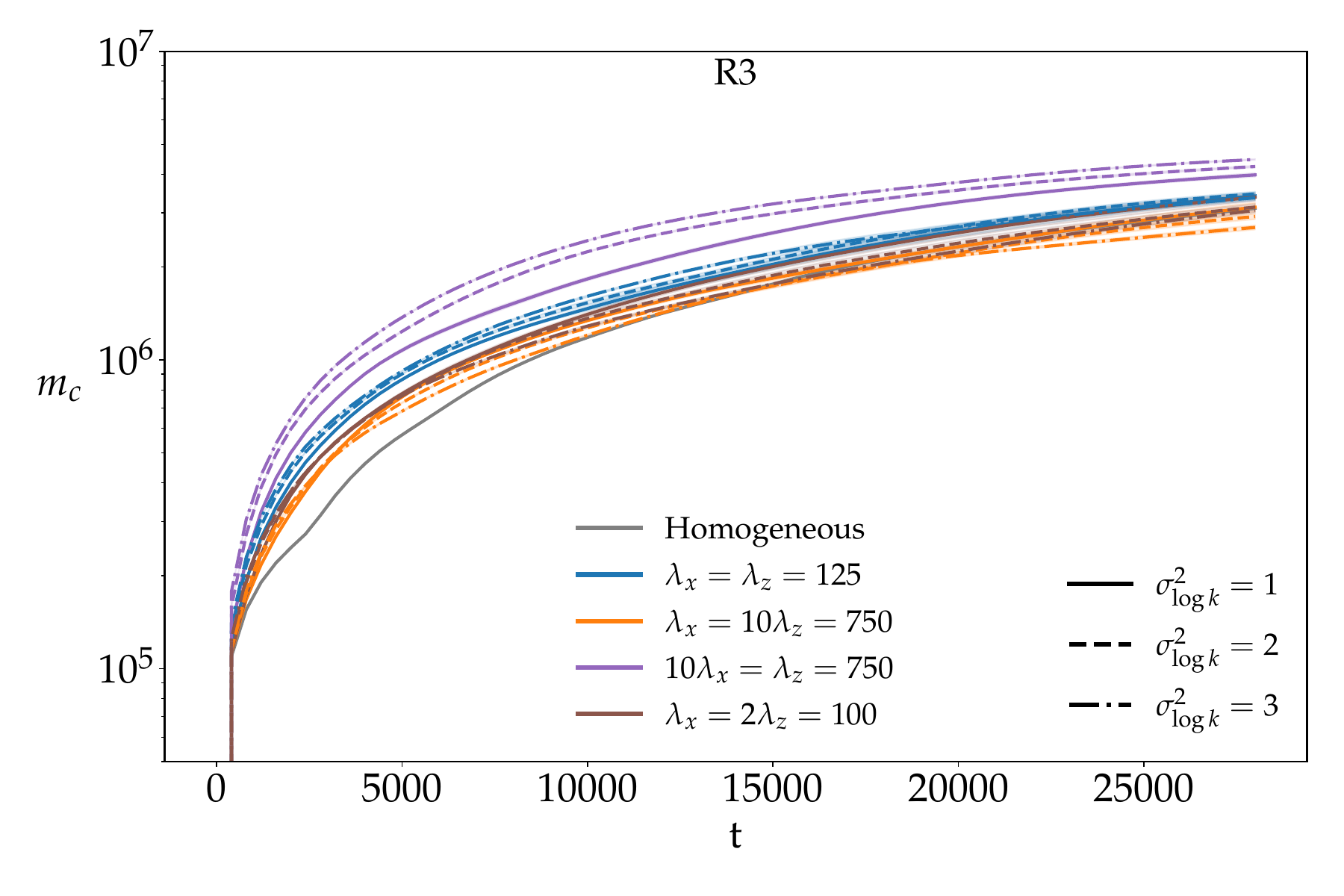}
    \caption{Amount of  product versus time for the homogeneous and multi-Gaussian heterogeneous cases. The figures present the same trend as in figure \ref{fig:mixlenlogs}. The shaded area shows the variability between realizations.}
    \label{fig:mclogs}
\end{figure}

The total amount of product  $\C$ (figures \ref{fig:mcstratcomp} and \ref{fig:mclogs}) follows the same tendency as the mixing length. The vertically stratified cases always present the steepest and fastest $m_{\C}$ particularly for the case R2 for which the highest quantity of $\C$ is produced. This is consistent with the previous observation of R2 cases having the widest reaction fronts. For the multi-gaussian cases, the media with the largest \lx produce the lowest $m_{\C}$, in agreement with the observed low reaction rate. The strength of the heterogeneity has the same effect as in the front width, when stratification is vertical, $m_{\C}$ is proportional to \slk{} and it is inversely proportional to \slk{} when the stratification is horizontal. In the multi-Gaussian cases (Figure \ref{fig:mclogs}), the effect is visible for the highest and lowest anisotropy ratios considered ($\lx = 10\lz=750$ and $10\lx = \lz=750$, respectively), while the rest of cases present intermediate behaviors. 

%
%
\subsubsection{Mixing}
Mixing in a reactive system can be analyzed by examining the conservative component $\cAC = \cA + \cC$. Component $\cAC$ follows an advection-diffusion equation that can be obtained by adding \eqref{eq:RDCA} and \eqref{eq:RDCC}. Therefore, mixing can be characterized by the scalar dissipation rate of $\cAC$ defined as
\begin{align}
  \label{eq:chi}
    \chi = \frac{1}{\Omega}\int_{\Omega} |\nabla (\cA + \cC)|^{2} d\Omega,
\end{align}
where $\Omega$ is the area of the domain.
\begin{figure}
    \centering
    \includegraphics[width=0.49\linewidth]{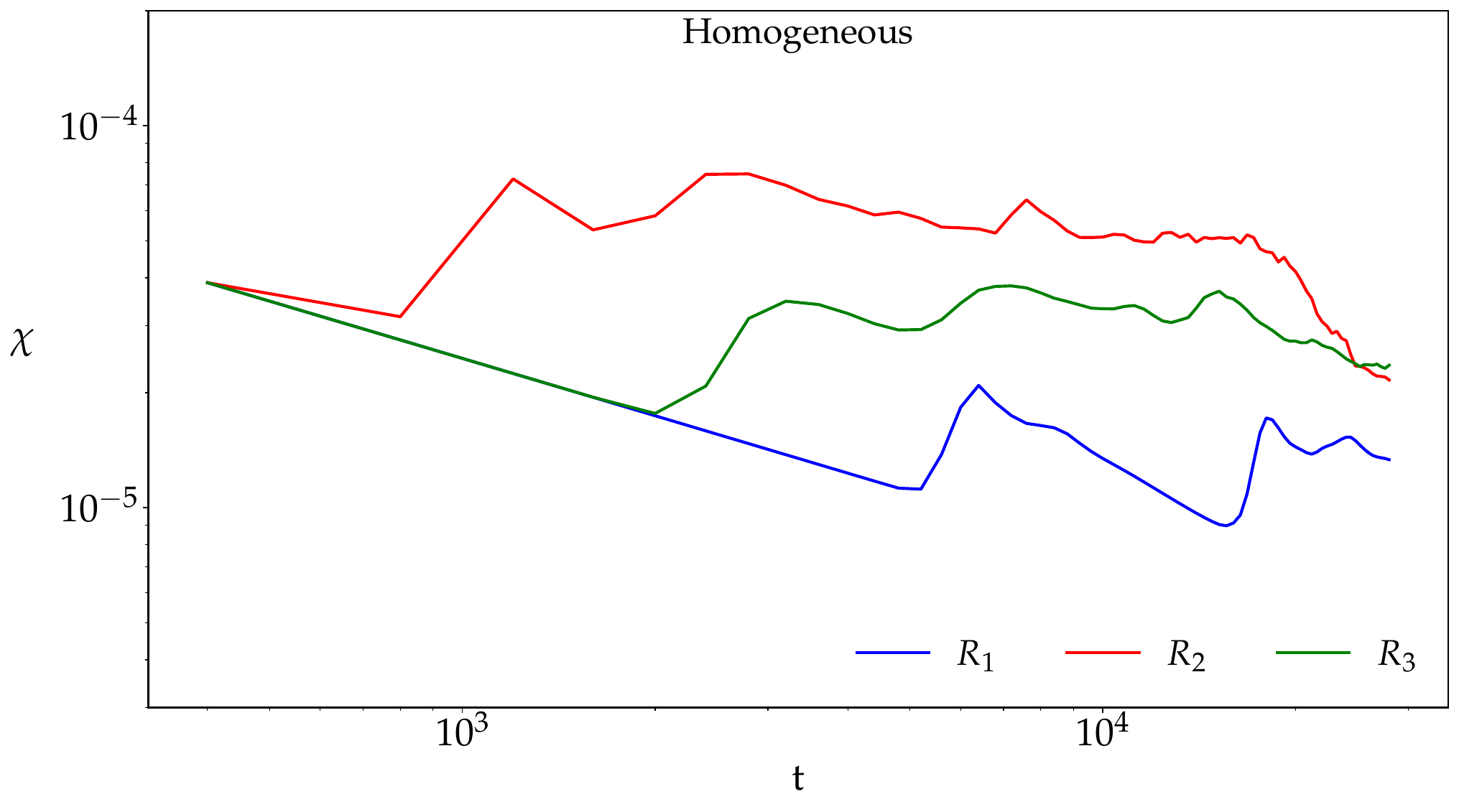}\\
    \includegraphics[width=0.49\linewidth]{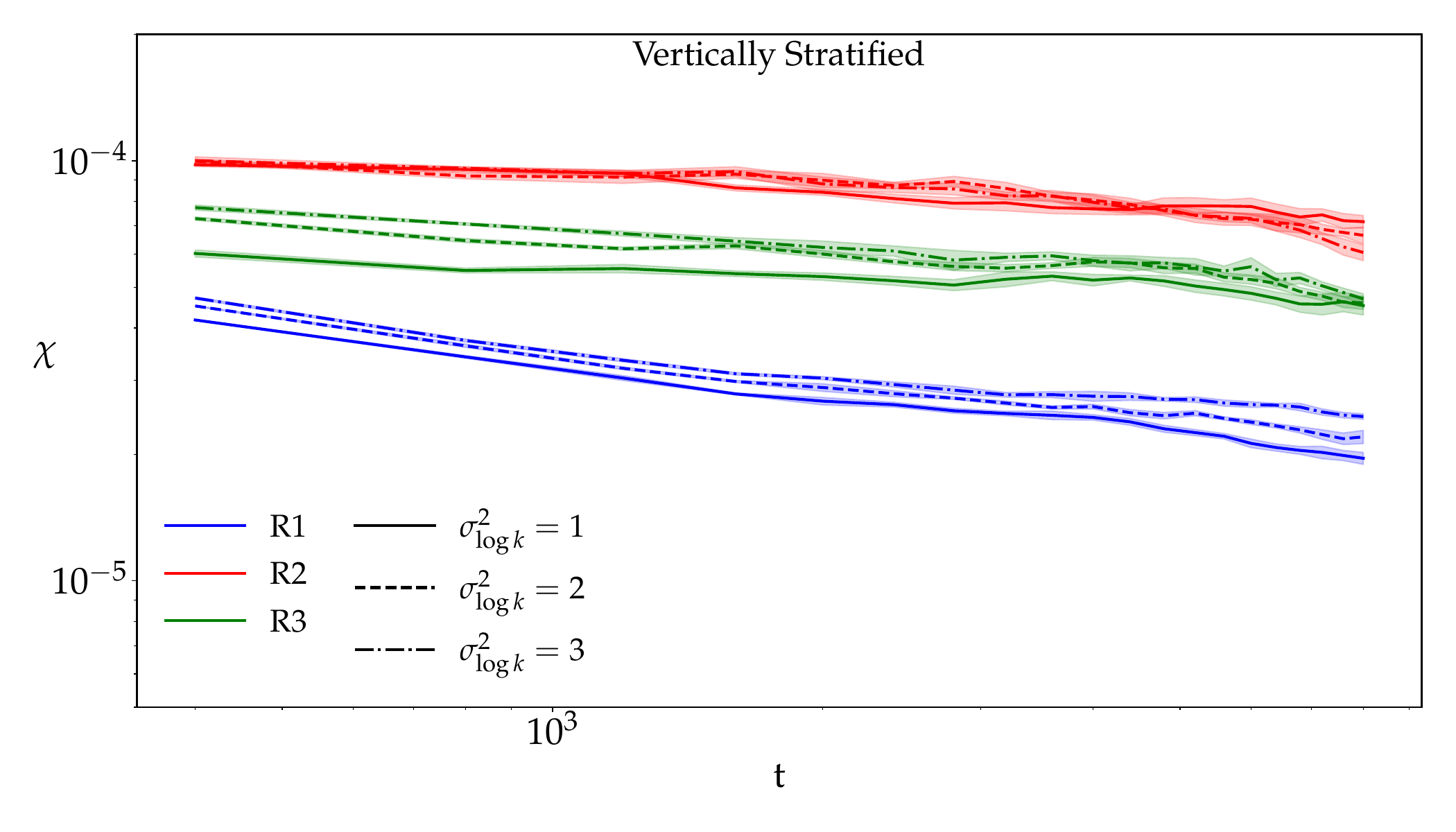}    \includegraphics[width=0.49\linewidth]{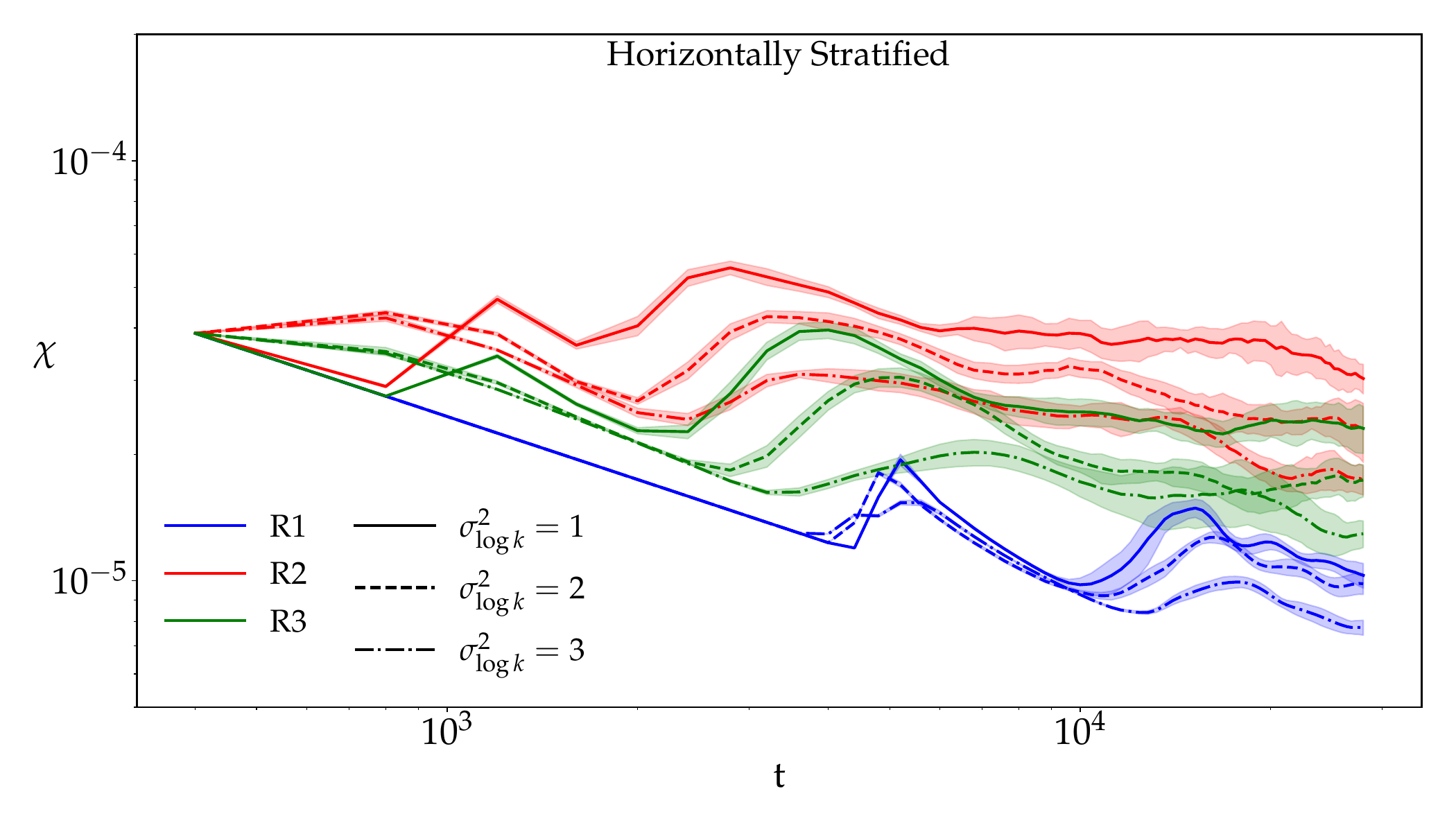}
    
    \caption{Scalar dissipation rate versus time for the homogeneous and all stratified cases. Horizontally stratified dissipation rates decrease with the variance whereas they increase with it in the vertically stratified one. Consistent with the trends observed for the previous parameters, R2 cases always present the highest values followed by R3 and R1. The shaded area shows the variability between realizations.}
    \label{fig:chivsts}
\end{figure}
\begin{figure}[htbp]
    \centering
    \includegraphics[width=\linewidth]{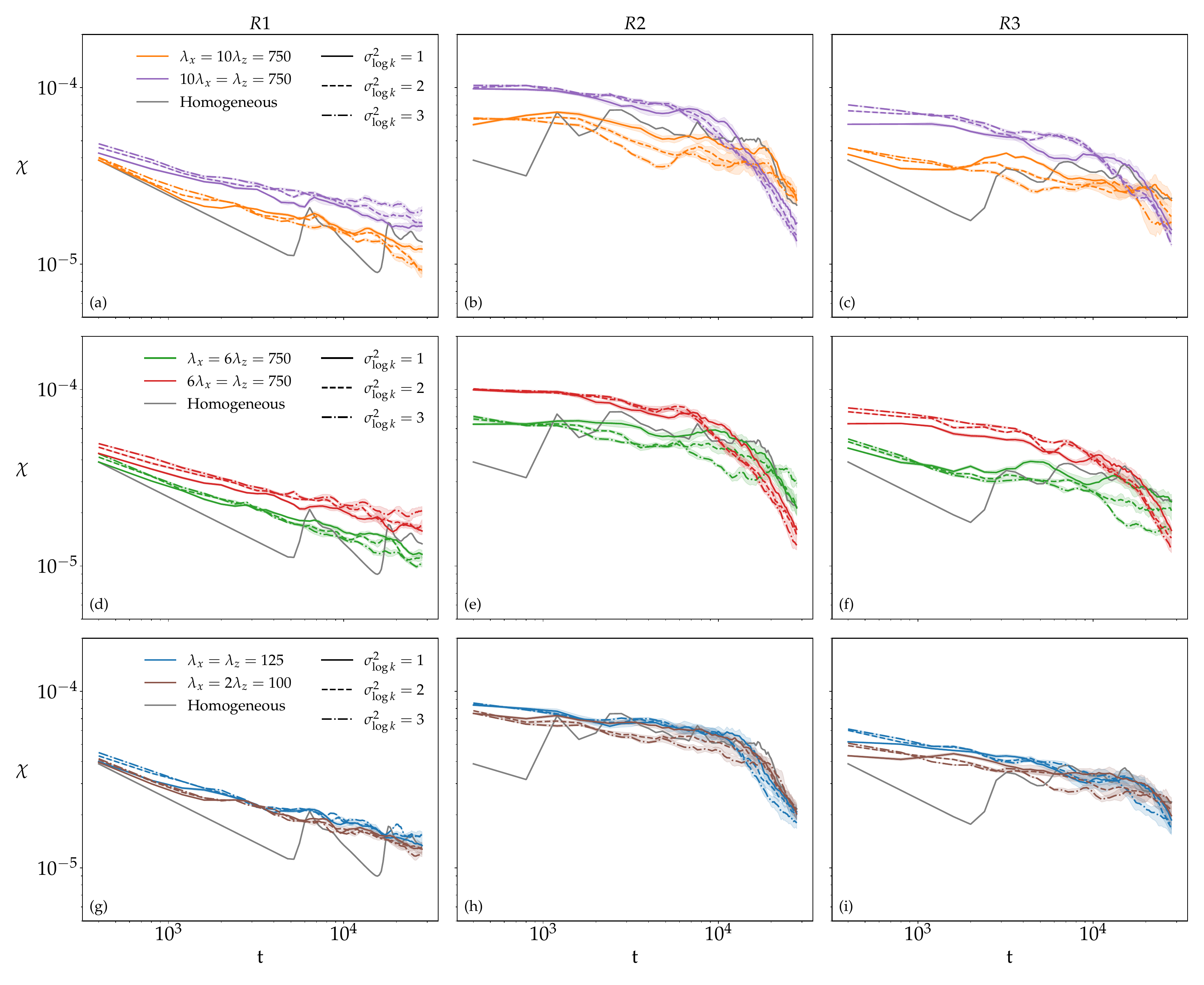}
    
    \caption{Scalar dissipation rate versus time for the homogeneous and multi-Gaussian heterogeneous cases. The scalar dissipation rate presents the same trend as the reaction rate in figure \ref{fig:rvstlog}. The shaded area shows the variability between realizations.}
    \label{fig:chivstlog}
\end{figure}

The scalar dissipation rates over the domain as a function of time (figures \ref{fig:chivsts} and \ref{fig:chivstlog}) exhibit a pattern almost identical to that of the reaction rate (see figures \ref{fig:rvsts} and \ref{fig:rvstlog}). Notably, the vertically stratified cases show the highest scalar dissipation and reaction rate, consistent with the rapid progression of the front position, mixing length and $m_{\C}$. In contrast, the horizontally stratified cases exhibit the lowest $\chi$. Overall, except for these horizontally stratified cases, which act as a barrier to flow, imposed heterogeneity enhances both reaction rates and scalar dissipation rates. Among all cases, R2 consistently exhibits the highest values for both quantities, followed by R3 and then R1. This ranking, previously observed for mixing lengths and $m_{\C}$, underscores the crucial role of bimolecular reactions in driving instability, on par with the impact of heterogeneity.  Moreover, increasing \slk{} decreases $\langle r \rangle$ and $\chi$ in the horizontally stratified cases whereas the opposite behavior is observed for the vertically stratified cases. As for the multi-Gaussian cases (figures \ref{fig:rvstlog} and \ref{fig:chivstlog}), the same tendency is observed as for $m_{\C}$ and the mixing length, depending on whether the anisotropy ration is greater or less than 1.
%
%
\section{Discussion and conclusions}
We numerically investigated the impact of heterogeneity on reactive  buoyancy-driven convective dissolution for the bimolecular reaction $\A + \B \rightarrow \C$, across various Rayleigh numbers. Our study explores homogeneous, horizontally stratified, vertically stratified, and multi-Gaussian permeability fields.  Our results show that key variables, including the total amount of product, mixing length, front position and width, reaction rate and scalar dissipation rate, are strongly influenced by heterogeneity. All these quantities reach larger values in vertically stratified and multi-Gaussian distributed permeability fields compared to homogeneous cases, which in turn sometimes exceed those in horizontally stratified fields.

The combined effect of heterogeneity and of the density stratification is responsible for the behavior of the different observables. In vertically stratified media, the alignment of the more permeable paths with gravity enhances mixing and finger growth, which in turn favors the chemical reaction. For all density cases, all observables grow as \slk{} increases with the exception of the reaction front velocity and width in the cases R2 and R3 (at very early times)  which  scale as $1/\slk$ (figures \ref{fig:frontposits} and \ref{fig:frontwmeans}). In these two cases, the reaction product contributes to the density of the fluid, therefore, the reaction increases the instability. As fingering becomes stronger, that is, \slk{} grows, the reaction product is carried away from the top boundary where most of the reaction happens. As a result, reaction in sinking fingers gets weaker as the heterogeneity strength increases and the growth of the reaction front velocity and width slows down.

In horizontally stratified media, all observables are proportional to $1/\slk$ for all cases. That is caused by the barrier effect of low permeable layers, that slow down the fingers, promote finger merging and hinder mixing and reaction. The exception is again the reaction front velocity and width, which are proportional to \slk{} for R2 and R3 (except at very early times, when the heterogenity has only been partially sampled and the behavior goes with \slk). R2 and R3 have less compact reaction fronts, therefore they are more sensitive to high permeability layers, that make the front advance faster and get wider as \slk{} increases.

The multi-Gaussian cases present a more intricate behavior. In most of the cases, no clear dependence on \slk{} is observed. Instead, the dynamics is dictated by the anisotropy ratio  $\lx/\lz$. When $\lx/\lz > 1$, resembling horizontal stratification, fingers propagate laterally with limited vertical progression. Conversely, when $\lx/\lz < 1$, mimicking vertical stratification, fingers advance more efficiently in the vertical direction with reduced wavelengths. Additionally, finger merging is more pronounced in cases with low anisotropy ratios. Moreover, the reaction rate and the amount of product grow as the anisotropy ratio decreases indicating more efficient mixing. As $\lx/\lz$ grows with large \lx, reaction rate, amount of product and the rest of the variables tend to behave as in the horizontally stratified media. In the opposite case (small $\lx/\lz$ and \lx), the behavior resembles that of the vertically stratified media.

In conclusion, the intensity and structure of heterogeneity control how flow, mixing and reaction scale with \slk. Moreover, the reaction affects density, which in turn alters buoyancy-driven convective flow, leading to opposing trends across stratified and multi-Gaussian fields.

These findings have direct implications for CO$_{2}$ sequestration, where selecting an optimal aquifer is crucial to minimizing leakage risks. Future work could extend this analysis to real porous media, such as those characterized via micro-tomography, and to other chemical reaction types as those inducing the dissolution of the porous medium \citep{hid15} or {\it in situ} precipitation \citep{moo19,Tho20}, to further refine our understanding of heterogeneity-driven convective processes.  

%
%
\begin{acknowledgments}
 RB, AD and JJH acknowledge the H2020 MSCA ITN program under the Grant No. 956457 (COPERMIX). RB and JJH acknowledge the support of the MICIU/AEI/10.13039/501100011033 and the European Union NextGenerationEU/PRTR through grant CNS2023-144134 (ESFERA).
\end{acknowledgments}
%
%
\bibliography{references.bib}

\begin{thebibliography}{34}%
\makeatletter
\providecommand \@ifxundefined [1]{%
 \@ifx{#1\undefined}
}%
\providecommand \@ifnum [1]{%
 \ifnum #1\expandafter \@firstoftwo
 \else \expandafter \@secondoftwo
 \fi
}%
\providecommand \@ifx [1]{%
 \ifx #1\expandafter \@firstoftwo
 \else \expandafter \@secondoftwo
 \fi
}%
\providecommand \natexlab [1]{#1}%
\providecommand \enquote  [1]{``#1''}%
\providecommand \bibnamefont  [1]{#1}%
\providecommand \bibfnamefont [1]{#1}%
\providecommand \citenamefont [1]{#1}%
\providecommand \href@noop [0]{\@secondoftwo}%
\providecommand \href [0]{\begingroup \@sanitize@url \@href}%
\providecommand \@href[1]{\@@startlink{#1}\@@href}%
\providecommand \@@href[1]{\endgroup#1\@@endlink}%
\providecommand \@sanitize@url [0]{\catcode `\\12\catcode `\$12\catcode
  `\&12\catcode `\#12\catcode `\^12\catcode `\_12\catcode `\%12\relax}%
\providecommand \@@startlink[1]{}%
\providecommand \@@endlink[0]{}%
\providecommand \url  [0]{\begingroup\@sanitize@url \@url }%
\providecommand \@url [1]{\endgroup\@href {#1}{\urlprefix }}%
\providecommand \urlprefix  [0]{URL }%
\providecommand \Eprint [0]{\href }%
\providecommand \doibase [0]{https://doi.org/}%
\providecommand \selectlanguage [0]{\@gobble}%
\providecommand \bibinfo  [0]{\@secondoftwo}%
\providecommand \bibfield  [0]{\@secondoftwo}%
\providecommand \translation [1]{[#1]}%
\providecommand \BibitemOpen [0]{}%
\providecommand \bibitemStop [0]{}%
\providecommand \bibitemNoStop [0]{.\EOS\space}%
\providecommand \EOS [0]{\spacefactor3000\relax}%
\providecommand \BibitemShut  [1]{\csname bibitem#1\endcsname}%
\let\auto@bib@innerbib\@empty
\bibitem [{\citenamefont {Kalam}\ \emph {et~al.}(2020)\citenamefont {Kalam},
  \citenamefont {Olayiwola}, \citenamefont {Al-Rubaii}, \citenamefont
  {Amaechi}, \citenamefont {Jamal},\ and\ \citenamefont
  {Awotunde}}]{Kalam2020}%
  \BibitemOpen
  \bibfield  {author} {\bibinfo {author} {\bibfnamefont {S.}~\bibnamefont
  {Kalam}}, \bibinfo {author} {\bibfnamefont {T.}~\bibnamefont {Olayiwola}},
  \bibinfo {author} {\bibfnamefont {M.~M.}\ \bibnamefont {Al-Rubaii}}, \bibinfo
  {author} {\bibfnamefont {B.~I.}\ \bibnamefont {Amaechi}}, \bibinfo {author}
  {\bibfnamefont {M.~S.}\ \bibnamefont {Jamal}},\ and\ \bibinfo {author}
  {\bibfnamefont {A.~A.}\ \bibnamefont {Awotunde}},\ }\bibfield  {title}
  {\bibinfo {title} {Carbon dioxide sequestration in underground formations:
  {R}eview of experimental, modeling, and field studies},\ }\href
  {https://doi.org/10.1007/s13202-020-01028-7} {\bibfield  {journal} {\bibinfo
  {journal} {J. Pet. Explor. Prod. Technol.}\ }\textbf {\bibinfo {volume}
  {11}},\ \bibinfo {pages} {303} (\bibinfo {year} {2020})}\BibitemShut
  {NoStop}%
\bibitem [{\citenamefont {Emami-Meybodi}\ \emph {et~al.}(2015)\citenamefont
  {Emami-Meybodi}, \citenamefont {Hassanzadeh}, \citenamefont {Green},\ and\
  \citenamefont {Ennis-King}}]{article_Emami}%
  \BibitemOpen
  \bibfield  {author} {\bibinfo {author} {\bibfnamefont {H.}~\bibnamefont
  {Emami-Meybodi}}, \bibinfo {author} {\bibfnamefont {H.}~\bibnamefont
  {Hassanzadeh}}, \bibinfo {author} {\bibfnamefont {C.}~\bibnamefont {Green}},\
  and\ \bibinfo {author} {\bibfnamefont {J.}~\bibnamefont {Ennis-King}},\
  }\bibfield  {title} {\bibinfo {title} {Convective dissolution of {CO$_{2}$}
  in saline aquifers: Progress in modeling and experiments},\ }\href
  {https://doi.org/10.1016/j.ijggc.2015.04.003} {\bibfield  {journal} {\bibinfo
   {journal} {Int. J. Greenhouse Gas Control}\ }\textbf {\bibinfo {volume}
  {40}},\ \bibinfo {pages} {238} (\bibinfo {year} {2015})}\BibitemShut
  {NoStop}%
\bibitem [{\citenamefont {Ching}\ \emph {et~al.}(2017)\citenamefont {Ching},
  \citenamefont {Chen},\ and\ \citenamefont {Tsai}}]{ching_convective_2017}%
  \BibitemOpen
  \bibfield  {author} {\bibinfo {author} {\bibfnamefont {J.}~\bibnamefont
  {Ching}}, \bibinfo {author} {\bibfnamefont {P.}~\bibnamefont {Chen}},\ and\
  \bibinfo {author} {\bibfnamefont {P.}~\bibnamefont {Tsai}},\ }\bibfield
  {title} {\bibinfo {title} {Convective mixing in homogeneous porous media
  flow},\ }\href {https://doi.org/10.1103/PhysRevFluids.2.014102} {\bibfield
  {journal} {\bibinfo  {journal} {Phys. Rev. Fluids}\ }\textbf {\bibinfo
  {volume} {2}},\ \bibinfo {pages} {014102} (\bibinfo {year}
  {2017})}\BibitemShut {NoStop}%
\bibitem [{\citenamefont {Hidalgo}\ \emph {et~al.}(2013)\citenamefont
  {Hidalgo}, \citenamefont {MacMinn},\ and\ \citenamefont {Juanes}}]{hid13}%
  \BibitemOpen
  \bibfield  {author} {\bibinfo {author} {\bibfnamefont {J.}~\bibnamefont
  {Hidalgo}}, \bibinfo {author} {\bibfnamefont {C.}~\bibnamefont {MacMinn}},\
  and\ \bibinfo {author} {\bibfnamefont {R.}~\bibnamefont {Juanes}},\
  }\bibfield  {title} {\bibinfo {title} {Dynamics of convective dissolution
  from a migrating current of carbon dioxide},\ }\href
  {https://doi.org/10.1016/j.advwatres.2013.06.013} {\bibfield  {journal}
  {\bibinfo  {journal} {Adv. Water Res.}\ }\textbf {\bibinfo {volume} {62}},\
  \bibinfo {pages} {511} (\bibinfo {year} {2013})}\BibitemShut {NoStop}%
\bibitem [{\citenamefont {Cheng}\ \emph {et~al.}(2013)\citenamefont {Cheng},
  \citenamefont {Lingzao},\ and\ \citenamefont {Liangsheng}}]{chen2013}%
  \BibitemOpen
  \bibfield  {author} {\bibinfo {author} {\bibfnamefont {C.}~\bibnamefont
  {Cheng}}, \bibinfo {author} {\bibfnamefont {Z.}~\bibnamefont {Lingzao}},\
  and\ \bibinfo {author} {\bibfnamefont {S.}~\bibnamefont {Liangsheng}},\
  }\bibfield  {title} {\bibinfo {title} {Continuum-scale convective mixing in
  geological {{CO$_{2}$}} sequestration in anisotropic and heterogeneous saline
  aquifers},\ }\href {https://doi.org/10.1016/j.advwatres.2012.10.012}
  {\bibfield  {journal} {\bibinfo  {journal} {Adv. Water Res.}\ }\textbf
  {\bibinfo {volume} {53}},\ \bibinfo {pages} {175} (\bibinfo {year}
  {2013})}\BibitemShut {NoStop}%
\bibitem [{\citenamefont {Green}\ and\ \citenamefont
  {Ennis-King}(2018)}]{article_Green}%
  \BibitemOpen
  \bibfield  {author} {\bibinfo {author} {\bibfnamefont {C.}~\bibnamefont
  {Green}}\ and\ \bibinfo {author} {\bibfnamefont {J.}~\bibnamefont
  {Ennis-King}},\ }\bibfield  {title} {\bibinfo {title} {Steady flux regime
  during convective mixing in three-dimensional heterogeneous porous media},\
  }\href {https://doi.org/10.3390/fluids3030058} {\bibfield  {journal}
  {\bibinfo  {journal} {Fluids}\ }\textbf {\bibinfo {volume} {3}},\ \bibinfo
  {pages} {58} (\bibinfo {year} {2018})}\BibitemShut {NoStop}%
\bibitem [{\citenamefont {Ghorbani}\ \emph {et~al.}(2017)\citenamefont
  {Ghorbani}, \citenamefont {Riaz},\ and\ \citenamefont
  {Daniel}}]{Ghorbani2017}%
  \BibitemOpen
  \bibfield  {author} {\bibinfo {author} {\bibfnamefont {Z.}~\bibnamefont
  {Ghorbani}}, \bibinfo {author} {\bibfnamefont {A.}~\bibnamefont {Riaz}},\
  and\ \bibinfo {author} {\bibfnamefont {D.}~\bibnamefont {Daniel}},\
  }\bibfield  {title} {\bibinfo {title} {Convective mixing in
  vertically-layered porous media: {T}he linear regime and the onset of
  convection},\ }\href {https://doi.org/10.1063/1.4996049} {\bibfield
  {journal} {\bibinfo  {journal} {Phys. Fluids}\ }\textbf {\bibinfo {volume}
  {29}},\ \bibinfo {pages} {084101} (\bibinfo {year} {2017})}\BibitemShut
  {NoStop}%
\bibitem [{\citenamefont {Farajzadeh}\ \emph {et~al.}(2011)\citenamefont
  {Farajzadeh}, \citenamefont {Ranganathan}, \citenamefont {Zitha},\ and\
  \citenamefont {Bruining}}]{article_Farajzadeh}%
  \BibitemOpen
  \bibfield  {author} {\bibinfo {author} {\bibfnamefont {R.}~\bibnamefont
  {Farajzadeh}}, \bibinfo {author} {\bibfnamefont {P.}~\bibnamefont
  {Ranganathan}}, \bibinfo {author} {\bibfnamefont {P.}~\bibnamefont {Zitha}},\
  and\ \bibinfo {author} {\bibfnamefont {J.}~\bibnamefont {Bruining}},\
  }\bibfield  {title} {\bibinfo {title} {The effect of heterogeneity on the
  character of density-driven natural convection of {CO$_{2}$} overlying a
  brine layer},\ }\href {https://doi.org/10.1016/j.advwatres.2010.12.012}
  {\bibfield  {journal} {\bibinfo  {journal} {Adv. Water Res.}\ }\textbf
  {\bibinfo {volume} {34}},\ \bibinfo {pages} {327} (\bibinfo {year}
  {2011})}\BibitemShut {NoStop}%
\bibitem [{\citenamefont {Ranganathan}\ \emph {et~al.}(2012)\citenamefont
  {Ranganathan}, \citenamefont {Farajzadeh}, \citenamefont {Bruining},\ and\
  \citenamefont {Zitha}}]{ranganathan2012}%
  \BibitemOpen
  \bibfield  {author} {\bibinfo {author} {\bibfnamefont {P.}~\bibnamefont
  {Ranganathan}}, \bibinfo {author} {\bibfnamefont {R.}~\bibnamefont
  {Farajzadeh}}, \bibinfo {author} {\bibfnamefont {H.}~\bibnamefont
  {Bruining}},\ and\ \bibinfo {author} {\bibfnamefont {P.~L.~J.}\ \bibnamefont
  {Zitha}},\ }\bibfield  {title} {\bibinfo {title} {Numerical simulation of
  natural convection in heterogeneous porous media for {CO$_{2}$} geological
  storage},\ }\href {https://doi.org/10.1007/s11242-012-0031-z} {\bibfield
  {journal} {\bibinfo  {journal} {Transp. Porous Media}\ }\textbf {\bibinfo
  {volume} {95}},\ \bibinfo {pages} {25} (\bibinfo {year} {2012})}\BibitemShut
  {NoStop}%
\bibitem [{\citenamefont {Qian}\ \emph {et~al.}(2019)\citenamefont {Qian},
  \citenamefont {Wei-Hua}, \citenamefont {Feng-Chen}, \citenamefont {Bingxi},\
  and\ \citenamefont {Ching-Yao}}]{Li2019}%
  \BibitemOpen
  \bibfield  {author} {\bibinfo {author} {\bibfnamefont {L.}~\bibnamefont
  {Qian}}, \bibinfo {author} {\bibfnamefont {C.}~\bibnamefont {Wei-Hua}},
  \bibinfo {author} {\bibfnamefont {L.}~\bibnamefont {Feng-Chen}}, \bibinfo
  {author} {\bibfnamefont {L.}~\bibnamefont {Bingxi}},\ and\ \bibinfo {author}
  {\bibfnamefont {C.}~\bibnamefont {Ching-Yao}},\ }\bibfield  {title} {\bibinfo
  {title} {Miscible density-driven flows in heterogeneous porous media:
  {I}nfluences of correlation length and distribution of permeability},\ }\href
  {https://doi.org/10.1103/PhysRevFluids.4.014502} {\bibfield  {journal}
  {\bibinfo  {journal} {Phys. Rev. Fluids}\ }\textbf {\bibinfo {volume} {4}},\
  \bibinfo {pages} {014502} (\bibinfo {year} {2019})}\BibitemShut {NoStop}%
\bibitem [{\citenamefont {Benhammadi}\ \emph {et~al.}(2025)\citenamefont
  {Benhammadi}, \citenamefont {Meunier},\ and\ \citenamefont
  {Hidalgo}}]{Benhammadi2025}%
  \BibitemOpen
  \bibfield  {author} {\bibinfo {author} {\bibfnamefont {R.}~\bibnamefont
  {Benhammadi}}, \bibinfo {author} {\bibfnamefont {P.}~\bibnamefont
  {Meunier}},\ and\ \bibinfo {author} {\bibfnamefont {J.~J.}\ \bibnamefont
  {Hidalgo}},\ }\bibfield  {title} {\bibinfo {title} {Experimental and
  numerical study of {CO$_2$} dissolution in a heterogeneous {H}ele-{S}haw
  cell},\ }\href {https://doi.org/10.1103/PhysRevFluids.10.043501} {\bibfield
  {journal} {\bibinfo  {journal} {Phys. Rev. Fluids}\ }\textbf {\bibinfo
  {volume} {10}},\ \bibinfo {pages} {043501} (\bibinfo {year}
  {2025})}\BibitemShut {NoStop}%
\bibitem [{\citenamefont {Brouzet}\ \emph {et~al.}(2022)\citenamefont
  {Brouzet}, \citenamefont {Méheust},\ and\ \citenamefont
  {Meunier}}]{article_Brouzet}%
  \BibitemOpen
  \bibfield  {author} {\bibinfo {author} {\bibfnamefont {C.}~\bibnamefont
  {Brouzet}}, \bibinfo {author} {\bibfnamefont {Y.}~\bibnamefont {Méheust}},\
  and\ \bibinfo {author} {\bibfnamefont {P.}~\bibnamefont {Meunier}},\
  }\bibfield  {title} {\bibinfo {title} {{{CO$_{2}$}} convective dissolution in
  a three-dimensional granular porous medium: {A}n experimental study},\ }\href
  {https://doi.org/10.1103/PhysRevFluids.7.033802} {\bibfield  {journal}
  {\bibinfo  {journal} {Phys. Rev. Fluids}\ }\textbf {\bibinfo {volume} {7}},\
  \bibinfo {pages} {033802} (\bibinfo {year} {2022})}\BibitemShut {NoStop}%
\bibitem [{\citenamefont {Hewitt}(2022)}]{Hewitt_2022}%
  \BibitemOpen
  \bibfield  {author} {\bibinfo {author} {\bibfnamefont {D.~R.}\ \bibnamefont
  {Hewitt}},\ }\bibfield  {title} {\bibinfo {title} {Evolution of convection in
  a layered porous medium},\ }\href {https://doi.org/10.1017/jfm.2022.335}
  {\bibfield  {journal} {\bibinfo  {journal} {J. Fluid Mech.}\ }\textbf
  {\bibinfo {volume} {941}},\ \bibinfo {pages} {A56} (\bibinfo {year}
  {2022})}\BibitemShut {NoStop}%
\bibitem [{\citenamefont {Almarcha}\ \emph {et~al.}(2010)\citenamefont
  {Almarcha}, \citenamefont {Trevelyan}, \citenamefont {Grosfils},\ and\
  \citenamefont {De~Wit}}]{Almarcha2009}%
  \BibitemOpen
  \bibfield  {author} {\bibinfo {author} {\bibfnamefont {C.}~\bibnamefont
  {Almarcha}}, \bibinfo {author} {\bibfnamefont {P.~M.~J.}\ \bibnamefont
  {Trevelyan}}, \bibinfo {author} {\bibfnamefont {P.}~\bibnamefont
  {Grosfils}},\ and\ \bibinfo {author} {\bibfnamefont {A.}~\bibnamefont
  {De~Wit}},\ }\bibfield  {title} {\bibinfo {title} {Chemically driven
  hydrodynamic instabilities},\ }\href
  {https://doi.org/10.1103/PhysRevLett.104.044501} {\bibfield  {journal}
  {\bibinfo  {journal} {Phys. Rev. Lett.}\ }\textbf {\bibinfo {volume} {104}},\
  \bibinfo {pages} {044501} (\bibinfo {year} {2010})}\BibitemShut {NoStop}%
\bibitem [{\citenamefont {Ghesmat}\ \emph {et~al.}(2011)\citenamefont
  {Ghesmat}, \citenamefont {Hassanzadeh},\ and\ \citenamefont {Abedi}}]{ghe11}%
  \BibitemOpen
  \bibfield  {author} {\bibinfo {author} {\bibfnamefont {K.}~\bibnamefont
  {Ghesmat}}, \bibinfo {author} {\bibfnamefont {H.}~\bibnamefont
  {Hassanzadeh}},\ and\ \bibinfo {author} {\bibfnamefont {J.}~\bibnamefont
  {Abedi}},\ }\bibfield  {title} {\bibinfo {title} {The impact of geochemistry
  on convective mixing in a gravitationally unstable diffusive boundary layer
  in porous media: {CO}$_2$ storage in saline aquifers},\ }\href
  {https://doi.org/10.1017/S0022112010006282} {\bibfield  {journal} {\bibinfo
  {journal} {J. Fluid Mech.}\ }\textbf {\bibinfo {volume} {673}},\ \bibinfo
  {pages} {480} (\bibinfo {year} {2011})}\BibitemShut {NoStop}%
\bibitem [{\citenamefont {Loodts}\ \emph {et~al.}(2014)\citenamefont {Loodts},
  \citenamefont {Thomas}, \citenamefont {Rongy},\ and\ \citenamefont
  {De~Wit}}]{Loodts2014}%
  \BibitemOpen
  \bibfield  {author} {\bibinfo {author} {\bibfnamefont {V.}~\bibnamefont
  {Loodts}}, \bibinfo {author} {\bibfnamefont {C.}~\bibnamefont {Thomas}},
  \bibinfo {author} {\bibfnamefont {L.}~\bibnamefont {Rongy}},\ and\ \bibinfo
  {author} {\bibfnamefont {A.}~\bibnamefont {De~Wit}},\ }\bibfield  {title}
  {\bibinfo {title} {Control of convective dissolution by chemical reactions:
  {G}eneral classification and application to {CO$_{2}$} dissolution in
  reactive aqueous solutions},\ }\href
  {https://doi.org/10.1103/PhysRevLett.113.114501} {\bibfield  {journal}
  {\bibinfo  {journal} {Phys. Rev. Lett.}\ }\textbf {\bibinfo {volume} {113}},\
  \bibinfo {pages} {114501} (\bibinfo {year} {2014})}\BibitemShut {NoStop}%
\bibitem [{\citenamefont {Cardoso}\ and\ \citenamefont {Andres}(2014)}]{car14}%
  \BibitemOpen
  \bibfield  {author} {\bibinfo {author} {\bibfnamefont {S.~S.~S.}\
  \bibnamefont {Cardoso}}\ and\ \bibinfo {author} {\bibfnamefont
  {J.}~\bibnamefont {Andres}},\ }\bibfield  {title} {\bibinfo {title}
  {Geochemistry of silicate-rich rocks can curtail spreading of carbon dioxide
  in subsurface aquifers},\ }\href {https://doi.org/10.1038/ncomms6743}
  {\bibfield  {journal} {\bibinfo  {journal} {Nat. Commun.}\ }\textbf {\bibinfo
  {volume} {5}},\ \bibinfo {pages} {5743} (\bibinfo {year} {2014})}\BibitemShut
  {NoStop}%
\bibitem [{\citenamefont {Loodts}\ \emph {et~al.}(2016)\citenamefont {Loodts},
  \citenamefont {Trevelyan}, \citenamefont {Rongy},\ and\ \citenamefont
  {De~Wit}}]{Loodts2016}%
  \BibitemOpen
  \bibfield  {author} {\bibinfo {author} {\bibfnamefont {V.}~\bibnamefont
  {Loodts}}, \bibinfo {author} {\bibfnamefont {P.}~\bibnamefont {Trevelyan}},
  \bibinfo {author} {\bibfnamefont {L.}~\bibnamefont {Rongy}},\ and\ \bibinfo
  {author} {\bibfnamefont {A.}~\bibnamefont {De~Wit}},\ }\bibfield  {title}
  {\bibinfo {title} {Density profiles around {A+B} $\to$ {C} reaction-diffusion
  fronts in partially miscible systems: {A} general classification},\ }\href
  {https://doi.org/10.1103/PhysRevE.94.043115} {\bibfield  {journal} {\bibinfo
  {journal} {Phys. Rev. E}\ }\textbf {\bibinfo {volume} {94}},\ \bibinfo
  {pages} {043115} (\bibinfo {year} {2016})}\BibitemShut {NoStop}%
\bibitem [{\citenamefont {Loodts}\ \emph {et~al.}(2017)\citenamefont {Loodts},
  \citenamefont {Knaepen}, \citenamefont {Rongy},\ and\ \citenamefont
  {De~Wit}}]{Loodts2017}%
  \BibitemOpen
  \bibfield  {author} {\bibinfo {author} {\bibfnamefont {V.}~\bibnamefont
  {Loodts}}, \bibinfo {author} {\bibfnamefont {B.}~\bibnamefont {Knaepen}},
  \bibinfo {author} {\bibfnamefont {L.}~\bibnamefont {Rongy}},\ and\ \bibinfo
  {author} {\bibfnamefont {A.}~\bibnamefont {De~Wit}},\ }\bibfield  {title}
  {\bibinfo {title} {Enhanced steady-state dissolution flux in reactive
  convective dissolution},\ }\href {https://doi.org/10.1039/C7CP01372H}
  {\bibfield  {journal} {\bibinfo  {journal} {Phys. Chem. Chem. Phys.}\
  }\textbf {\bibinfo {volume} {19}},\ \bibinfo {pages} {18565} (\bibinfo {year}
  {2017})}\BibitemShut {NoStop}%
\bibitem [{\citenamefont {Ghoshal}\ and\ \citenamefont
  {Cardoso}(2018)}]{Ghoshal2018}%
  \BibitemOpen
  \bibfield  {author} {\bibinfo {author} {\bibfnamefont {P.}~\bibnamefont
  {Ghoshal}}\ and\ \bibinfo {author} {\bibfnamefont {S.~S.~S.}\ \bibnamefont
  {Cardoso}},\ }\bibfield  {title} {\bibinfo {title} {Reactive
  convective-dissolution in a porous medium: stability and nonlinear
  dynamics},\ }\href {https://doi.org/10.1039/c8cp03064b} {\bibfield  {journal}
  {\bibinfo  {journal} {Phys. Chem. Chem. Phys.}\ }\textbf {\bibinfo {volume}
  {20}},\ \bibinfo {pages} {21617} (\bibinfo {year} {2018})}\BibitemShut
  {NoStop}%
\bibitem [{\citenamefont {Jotkar}\ \emph {et~al.}(2019)\citenamefont {Jotkar},
  \citenamefont {Rongy},\ and\ \citenamefont {De~Wit}}]{Jotkar2019b}%
  \BibitemOpen
  \bibfield  {author} {\bibinfo {author} {\bibfnamefont {M.}~\bibnamefont
  {Jotkar}}, \bibinfo {author} {\bibfnamefont {L.}~\bibnamefont {Rongy}},\ and\
  \bibinfo {author} {\bibfnamefont {A.}~\bibnamefont {De~Wit}},\ }\bibfield
  {title} {\bibinfo {title} {Chemically-driven convective dissolution},\ }\href
  {https://doi.org/10.1039/c9cp03044a} {\bibfield  {journal} {\bibinfo
  {journal} {Phys. Chem. Chem. Phys.}\ }\textbf {\bibinfo {volume} {21}},\
  \bibinfo {pages} {19054} (\bibinfo {year} {2019})}\BibitemShut {NoStop}%
\bibitem [{\citenamefont {Lei}\ and\ \citenamefont {Luo}(2019)}]{lei19}%
  \BibitemOpen
  \bibfield  {author} {\bibinfo {author} {\bibfnamefont {T.}~\bibnamefont
  {Lei}}\ and\ \bibinfo {author} {\bibfnamefont {K.}~\bibnamefont {Luo}},\
  }\bibfield  {title} {\bibinfo {title} {Pore-scale study of dissolution-driven
  density instability with reaction {A+B} $\to$ {C} in porous media},\ }\href
  {https://doi.org/10.1103/PhysRevFluids.4.063907} {\bibfield  {journal}
  {\bibinfo  {journal} {Phys. Rev. Fluids}\ }\textbf {\bibinfo {volume} {4}},\
  \bibinfo {pages} {063907} (\bibinfo {year} {2019})}\BibitemShut {NoStop}%
\bibitem [{\citenamefont {Thomas}\ \emph {et~al.}(2020)\citenamefont {Thomas},
  \citenamefont {Dehaeck},\ and\ \citenamefont {De~Wit}}]{Tho20}%
  \BibitemOpen
  \bibfield  {author} {\bibinfo {author} {\bibfnamefont {C.}~\bibnamefont
  {Thomas}}, \bibinfo {author} {\bibfnamefont {S.}~\bibnamefont {Dehaeck}},\
  and\ \bibinfo {author} {\bibfnamefont {A.}~\bibnamefont {De~Wit}},\
  }\bibfield  {title} {\bibinfo {title} {Effect of precipitation mineralization
  reactions on convective dissolution of {CO$_{2}$}: {A}n experimental study},\
  }\bibfield  {journal} {\bibinfo  {journal} {Phys. Rev. Fluids}\ }\textbf
  {\bibinfo {volume} {5}},\ \href
  {https://doi.org/10.1103/physrevfluids.5.113505}
  {10.1103/physrevfluids.5.113505} (\bibinfo {year} {2020})\BibitemShut
  {NoStop}%
\bibitem [{\citenamefont {Jotkar}\ \emph {et~al.}(2020)\citenamefont {Jotkar},
  \citenamefont {Rongy},\ and\ \citenamefont {De~Wit}}]{Jotkar2020}%
  \BibitemOpen
  \bibfield  {author} {\bibinfo {author} {\bibfnamefont {M.}~\bibnamefont
  {Jotkar}}, \bibinfo {author} {\bibfnamefont {L.}~\bibnamefont {Rongy}},\ and\
  \bibinfo {author} {\bibfnamefont {A.}~\bibnamefont {De~Wit}},\ }\bibfield
  {title} {\bibinfo {title} {Reactive convective dissolution with differential
  diffusivities: {N}onlinear simulations of onset times and asymptotic
  fluxes},\ }\href {https://doi.org/10.1103/PhysRevFluids.5.104502} {\bibfield
  {journal} {\bibinfo  {journal} {Phys. Rev. Fluids}\ }\textbf {\bibinfo
  {volume} {5}},\ \bibinfo {pages} {104502} (\bibinfo {year}
  {2020})}\BibitemShut {NoStop}%
\bibitem [{\citenamefont {Jotkar}\ \emph {et~al.}(2021)\citenamefont {Jotkar},
  \citenamefont {De~Wit},\ and\ \citenamefont {Rongy}}]{Jotkar2021}%
  \BibitemOpen
  \bibfield  {author} {\bibinfo {author} {\bibfnamefont {M.}~\bibnamefont
  {Jotkar}}, \bibinfo {author} {\bibfnamefont {A.}~\bibnamefont {De~Wit}},\
  and\ \bibinfo {author} {\bibfnamefont {L.}~\bibnamefont {Rongy}},\ }\bibfield
   {title} {\bibinfo {title} {Control of chemically driven convective
  dissolution by differential diffusion effects},\ }\href
  {https://doi.org/10.5194/egusphere-egu21-8718} {\bibfield  {journal}
  {\bibinfo  {journal} {Phys. Rev. Fluids}\ }\textbf {\bibinfo {volume} {6}},\
  \bibinfo {pages} {053504} (\bibinfo {year} {2021})}\BibitemShut {NoStop}%
\bibitem [{\citenamefont {Loodts}\ \emph {et~al.}(2015)\citenamefont {Loodts},
  \citenamefont {Rongy},\ and\ \citenamefont {De~Wit}}]{Loodts2015}%
  \BibitemOpen
  \bibfield  {author} {\bibinfo {author} {\bibfnamefont {V.}~\bibnamefont
  {Loodts}}, \bibinfo {author} {\bibfnamefont {L.}~\bibnamefont {Rongy}},\ and\
  \bibinfo {author} {\bibfnamefont {A.}~\bibnamefont {De~Wit}},\ }\bibfield
  {title} {\bibinfo {title} {Chemical control of dissolution-driven convection
  in partially miscible systems: theoretical classification},\ }\href
  {https://doi.org/10.1039/C5CP03082J} {\bibfield  {journal} {\bibinfo
  {journal} {Phys. Chem. Chem. Phys.}\ }\textbf {\bibinfo {volume} {17}},\
  \bibinfo {pages} {29814} (\bibinfo {year} {2015})}\BibitemShut {NoStop}%
\bibitem [{\citenamefont {Icardi}\ \emph {et~al.}(2023)\citenamefont {Icardi},
  \citenamefont {Pescimoro}, \citenamefont {Municchi},\ and\ \citenamefont
  {Hidalgo}}]{Icardi2023}%
  \BibitemOpen
  \bibfield  {author} {\bibinfo {author} {\bibfnamefont {M.}~\bibnamefont
  {Icardi}}, \bibinfo {author} {\bibfnamefont {E.}~\bibnamefont {Pescimoro}},
  \bibinfo {author} {\bibfnamefont {F.}~\bibnamefont {Municchi}},\ and\
  \bibinfo {author} {\bibfnamefont {J.~J.}\ \bibnamefont {Hidalgo}},\
  }\bibfield  {title} {\bibinfo {title} {Computational framework for complex
  flow and transport in heterogeneous porous media},\ }\href
  {https://doi.org/10.1007/s00366-023-01825-8} {\bibfield  {journal} {\bibinfo
  {journal} {Eng. Comput.}\ }\textbf {\bibinfo {volume} {39}},\ \bibinfo
  {pages} {1} (\bibinfo {year} {2023})}\BibitemShut {NoStop}%
\bibitem [{\citenamefont {Weller}\ \emph {et~al.}(1998)\citenamefont {Weller},
  \citenamefont {Tabor}, \citenamefont {Jasak},\ and\ \citenamefont
  {Fureby}}]{Weller1998}%
  \BibitemOpen
  \bibfield  {author} {\bibinfo {author} {\bibfnamefont {H.~G.}\ \bibnamefont
  {Weller}}, \bibinfo {author} {\bibfnamefont {G.}~\bibnamefont {Tabor}},
  \bibinfo {author} {\bibfnamefont {H.}~\bibnamefont {Jasak}},\ and\ \bibinfo
  {author} {\bibfnamefont {C.}~\bibnamefont {Fureby}},\ }\bibfield  {title}
  {\bibinfo {title} {A tensorial approach to computational continuum mechanics
  using object-oriented techniques},\ }\href {https://doi.org/10.1063/1.168744}
  {\bibfield  {journal} {\bibinfo  {journal} {CIP}\ }\textbf {\bibinfo {volume}
  {12}},\ \bibinfo {pages} {620} (\bibinfo {year} {1998})}\BibitemShut
  {NoStop}%
\bibitem [{\citenamefont {De~Wit}(2004)}]{Wit2004}%
  \BibitemOpen
  \bibfield  {author} {\bibinfo {author} {\bibfnamefont {A.}~\bibnamefont
  {De~Wit}},\ }\bibfield  {title} {\bibinfo {title} {Miscible density fingering
  of chemical fronts in porous media: {N}onlinear simulations},\ }\href
  {https://doi.org/10.1063/1.1630576} {\bibfield  {journal} {\bibinfo
  {journal} {Phys. Fluids}\ }\textbf {\bibinfo {volume} {16}},\ \bibinfo
  {pages} {163} (\bibinfo {year} {2004})}\BibitemShut {NoStop}%
\bibitem [{\citenamefont {De~Wit}\ and\ \citenamefont
  {Homsy}(1997)}]{DeWit1997b}%
  \BibitemOpen
  \bibfield  {author} {\bibinfo {author} {\bibfnamefont {A.}~\bibnamefont
  {De~Wit}}\ and\ \bibinfo {author} {\bibfnamefont {G.~M.}\ \bibnamefont
  {Homsy}},\ }\bibfield  {title} {\bibinfo {title} {Viscous fingering in
  periodically heterogeneous porous media. {II.} {N}umerical simulations},\
  }\href {https://doi.org/10.1063/1.475259} {\bibfield  {journal} {\bibinfo
  {journal} {J. Chem. Phys.}\ }\textbf {\bibinfo {volume} {107}},\ \bibinfo
  {pages} {9619} (\bibinfo {year} {1997})}\BibitemShut {NoStop}%
\bibitem [{\citenamefont {Gopalakrishnan}\ \emph {et~al.}(2017)\citenamefont
  {Gopalakrishnan}, \citenamefont {Carballido-Landeira}, \citenamefont
  {De~Wit},\ and\ \citenamefont {Knaepen}}]{Gopalakrishnan2017}%
  \BibitemOpen
  \bibfield  {author} {\bibinfo {author} {\bibfnamefont {S.~S.}\ \bibnamefont
  {Gopalakrishnan}}, \bibinfo {author} {\bibfnamefont {J.}~\bibnamefont
  {Carballido-Landeira}}, \bibinfo {author} {\bibfnamefont {A.}~\bibnamefont
  {De~Wit}},\ and\ \bibinfo {author} {\bibfnamefont {B.}~\bibnamefont
  {Knaepen}},\ }\bibfield  {title} {\bibinfo {title} {Relative role of
  convective and diffusive mixing in the miscible {R}ayleigh-{T}aylor
  instability in porous media},\ }\href
  {https://doi.org/10.1103/physrevfluids.2.012501} {\bibfield  {journal}
  {\bibinfo  {journal} {Phys. Rev. Fluids}\ }\textbf {\bibinfo {volume} {2}},\
  \bibinfo {pages} {012501(R)} (\bibinfo {year} {2017})}\BibitemShut {NoStop}%
\bibitem [{\citenamefont {Dentz}\ \emph {et~al.}(2022)\citenamefont {Dentz},
  \citenamefont {Hidalgo},\ and\ \citenamefont {Lester}}]{Dentz2022}%
  \BibitemOpen
  \bibfield  {author} {\bibinfo {author} {\bibfnamefont {M.}~\bibnamefont
  {Dentz}}, \bibinfo {author} {\bibfnamefont {J.~J.}\ \bibnamefont {Hidalgo}},\
  and\ \bibinfo {author} {\bibfnamefont {D.}~\bibnamefont {Lester}},\
  }\bibfield  {title} {\bibinfo {title} {Mixing in porous media: {C}oncepts and
  approaches across scales},\ }\href
  {https://doi.org/10.1007/s11242-022-01852-x} {\bibfield  {journal} {\bibinfo
  {journal} {Transp. Porous Media}\ }\textbf {\bibinfo {volume} {146}},\
  \bibinfo {pages} {5} (\bibinfo {year} {2022})}\BibitemShut {NoStop}%
\bibitem [{\citenamefont {Hidalgo}\ \emph {et~al.}(2015)\citenamefont
  {Hidalgo}, \citenamefont {Dentz}, \citenamefont {Cabeza},\ and\ \citenamefont
  {Carrera}}]{hid15}%
  \BibitemOpen
  \bibfield  {author} {\bibinfo {author} {\bibfnamefont {J.}~\bibnamefont
  {Hidalgo}}, \bibinfo {author} {\bibfnamefont {M.}~\bibnamefont {Dentz}},
  \bibinfo {author} {\bibfnamefont {Y.}~\bibnamefont {Cabeza}},\ and\ \bibinfo
  {author} {\bibfnamefont {J.}~\bibnamefont {Carrera}},\ }\bibfield  {title}
  {\bibinfo {title} {Dissolution patterns and mixing dynamics in unstable
  reactive flow},\ }\href {https://doi.org/10.1002/2015GL065036} {\bibfield
  {journal} {\bibinfo  {journal} {Geophys. Res. Lett.}\ }\textbf {\bibinfo
  {volume} {42}},\ \bibinfo {pages} {6357} (\bibinfo {year}
  {2015})}\BibitemShut {NoStop}%
\bibitem [{\citenamefont {Moosavi}\ \emph {et~al.}(2019)\citenamefont
  {Moosavi}, \citenamefont {Kumar}, \citenamefont {De~Wit},\ and\ \citenamefont
  {Schr\"oter}}]{moo19}%
  \BibitemOpen
  \bibfield  {author} {\bibinfo {author} {\bibfnamefont {R.}~\bibnamefont
  {Moosavi}}, \bibinfo {author} {\bibfnamefont {A.}~\bibnamefont {Kumar}},
  \bibinfo {author} {\bibfnamefont {A.}~\bibnamefont {De~Wit}},\ and\ \bibinfo
  {author} {\bibfnamefont {M.}~\bibnamefont {Schr\"oter}},\ }\bibfield  {title}
  {\bibinfo {title} {Influence of mineralization and injection flow rate on
  flow patterns in three-dimensional porous media},\ }\href
  {https://doi.org/10.1039/C9CP01382B} {\bibfield  {journal} {\bibinfo
  {journal} {Phys. Chem. Chem. Phys.}\ }\textbf {\bibinfo {volume} {21}},\
  \bibinfo {pages} {14605} (\bibinfo {year} {2019})}\BibitemShut {NoStop}%
\end{thebibliography}%

\end{document}